\documentclass[12pt]{article}

\usepackage{amssymb,amsmath,amsthm, amstext}
\usepackage{latexsym}
\usepackage{natbib}
\usepackage{amsbsy}
\usepackage{graphicx}
\usepackage{float}
\usepackage{subfloat}
\usepackage{caption}
\usepackage{subcaption}
\usepackage{epsfig}
\usepackage{epstopdf}  
\usepackage{color}
\usepackage[font= {small,it} ]{caption}
\usepackage{url}
\usepackage[titletoc]{appendix}
\usepackage{xcolor}  
\usepackage{hyperref}
\usepackage[margin= 1in]{geometry}
\usepackage{appendix}
\usepackage{bm}
\usepackage{multirow}

\usepackage[ruled,vlined]{algorithm2e}
\usepackage[noend]{algpseudocode}

\newcommand{\indep}{\rotatebox[origin=c]{90}{$\models$}}
\DeclareMathOperator*{\argmax}{arg\,max}
\DeclareMathOperator*{\argmin}{arg\,min}
\DeclareMathOperator{\tr}{tr}
\DeclareMathOperator{\cov}{cov}
\DeclareMathOperator{\var}{var}

\newcommand{\B}{\boldsymbol}
\newcommand*\samethanks[1][\value{footnote}]{\footnotemark[#1]}

\begin{document}

\title{Bayesian Regularization for Functional Graphical Models with Applications to Neuroimaging}
\author{Jiajing Niu\thanks{School of Mathematical and Statistical Sciences, Clemson University, Clemson, SC 29634, USA} \and Boyoung Hur\samethanks \and John Absher\thanks{Prisma Health, Greenville, SC, 29605}\and D. Andrew Brown\thanks{Corresponding Author, School of Mathematical and Statiscal Sciences, Clemson University, Clemson, SC 29634}\\[8pt] for the Alzheimer's Disease Neuroimaging Initiative\footnote{Data used in preparation of this article were obtained from the Alzheimer's Disease
Neuroimaging Initiative (ADNI) database (adni.loni.usc.edu). As such, the investigators
within the ADNI contributed to the design and implementation of ADNI and/or provided data
but did not participate in analysis or writing of this report. A complete listing of ADNI
investigators can be found at:
\url{http://adni.loni.usc.edu/wp-content/uploads/how_to_apply/ADNI_Acknowledgement_List.pdf}}}

\maketitle

\begin{abstract}
Graphical models, used to express conditional dependence between random variables observed at various nodes, are used extensively in many fields such as genetics, neuroscience, and social network analysis. While most current statistical methods for estimating graphical models focus on scalar data, there is interest in estimating analogous dependence structures when the data observed at each node are functional, such as signals or images. In this paper, we propose a fully Bayesian regularization scheme for estimating functional graphical models. We first consider a direct Bayesian analog of the functional graphical lasso proposed by \cite{qiao2019functional}. We then propose a regularization strategy via the graphical horseshoe. We compare these approaches via simulation study and apply our proposed functional graphical horseshoe to two motivating applications, electroencephalography data for comparing brain activation between an alcoholic group and controls, as well as changes in structural connectivity in the presence of traumatic brain injury (TBI). Our results yield insight into how the brain attempts to compensate for disconnected networks after injury.

\end{abstract}

\section{Introduction}\label{sec:intro}
Graphical models use graphs to model and draw inferences concerning conditional independence among a collection of random variables or processes, each of which is associated with a particular location (also called a node or a vertex). They have been used to study flow cytometry between cell proteins \citep{friedman2008sparse}, to estimate networks from gene expression data \citep{li2019graphical}, and to identify communicating regions from electroencephalography (EEG) data \citep{qiao2019functional}. In this work we are focused on Gaussian graphical models, where the data follow a multivariate Gaussian distribution. In this case estimating the edge set is equivalent to identifying the nonzero elements of the precision matrix associated with the Gaussian distribution. 

Broadly speaking, frequentist studies of graphical models have either involved neighborhood selection \citep{meinshausen2006high} or the graphical lasso \citep{yuan2007model, friedman2008sparse}. The neighborhood selection method employs regression of each variable on the remaining variables with regularization, and then summarizing the neighborhoods together. On the other hand, \cite{friedman2008sparse} proposed the graphical lasso via a Gaussian log-likelihood with the lasso regularization on the entire precision matrix. The glasso has proven to be useful and is a widely used procedure, due to  the sparsity and convergence rates that have been studied \citep{lam2009sparsistency} as well as associated computational techniques \citep{friedman2008sparse, zhu2014structural}. A Bayesian version of the graphical lasso was proposed by \cite{wang2012bayesian}, who illustrated potential differences between the posterior mean and the posterior mode that might be encountered. \cite{li2019graphical} extended the ideas of \cite{wang2012bayesian} by proposing a graphical horseshoe estimator, along with an efficient Markov chain Monte Carlo (MCMC)\citep{GelfandSmith90} algorithm for its implementation. 

To date, most of the graphical modeling literature has focused on data in which each node has an associated scalar or vector-valued response variable. However, many real world applications involve the collection of functional data at each node. In this case, we have a collection of subjects / units for whom a set of continuously-supported random functions are (discretely) observed, one function at each node, where the support may be time- or spatially-indexed, or both. For example, in neuroscience there is much interest in studying connectivity; e.g., in terms of connected regions of interest measured in functional magnetic resonance imaging (fMRI)\citep{ShappelEtAl19} or communicating electrodes in electroencephalography (EEG)\citep{zhang1995event} corresponding to associated regions of neuronal activity. Alternatively, in social network analysis and marketing, it is possible to observe and record online behavior patterns among baskets of different goods for each customer over a period of time to identify related types of products. Compared to scalar or vector-valued graphical models, functional graphical models remain vastly underexplored. \cite{qiao2019functional} proposed a functional version of the graphical lasso along with a block-coordinate descent algorithm for optimizing the loss function. \cite{qiao2020doubly} model such data using doubly functional graphical models through a nonparametric approach to smooth $p$ covariance matrices, where the graph is functional in nature. Similarly, \cite{zapata2019partial} decomposed a functional graphical model into a sequence of standard multivariate graphical models under an assumption of partial separability for multivariate functional data. \cite{li2018nonparametric} proposed a nonparametric functional graphical model based on additive conditional dependence via nested Hilbert spaces and additive precision operators. Further, \cite{solea2020copula} relaxed the multivariate Gaussian process assumption by introducing the functional copula Gaussian graphical model. \cite{zhu2016bayesian} proposed a Bayesian framework for working with functional graphical models directly in the space of infinite-dimensional random functions, essentially extending the work of \cite{DawidLauritzen93} to function space by using hyper-inverse Wishart priors on the space of plausible, decomposable graphs. Recently, \cite{Zhang_etal_21} proposed a Bayesian model for functional graphical models in which independent Laplace priors are placed on reparameterized partial correlations associated with basis coefficients, inducing a so-called normal hypo-exponential shrinkage prior and allowing the graph to functionally evolve over time. They utilize basis function representations that model the within-functional correlations, then employ Bayesian regularization in the basis space assuming independence of basis coefficients across different nodes. Compared with this work, our proposed method is for a overall static graph with assumption that different basis coefficients could be dependent across nodes.\\ 
\hspace*{12pt}Our work is motivated by neuroimaging data that typically have low signal-to-noise ratios caused by non-neural noise arising from cardiac and respiratory processes or scanner instability, a problem that is exacerbated by the typically small numbers of subjects available from such studies. For instance, in Section \ref{sec:ADNI} we study the effects of traumatic brain injury on connectivity of the human brain. The diffusion-weighted magnetic resonance imaging data consist of longitudinal measurements of white matter integrity within 26 regions of interest in 34 subjects, 17 of whom have been diagnosed with a traumatic brain injury (TBI). We aim to assess chronic structural connectivity differences between the TBI and non-TBI groups using the \cite{GreenlawEtAl17} used an imaging genetics example to demonstrate dramatic differences in associations between genetic variations and brain imaging measures that might be identified when accounting for uncertainty in a model estimate versus using an optimization-based point estimate alone.

Functional graphical models have been proposed and studied assuming both static\citep{zhu2016bayesian, qiao2019functional} and dynamic underlying graphs\citep{warnick2018bayesian, Zhang_etal_21}. Much existing work on dynamic graphs is typically motivated by fMRI studies, since some evidence suggests that functional activity behaves in accordance to networks that move between a finite number of states over the course of an experiment\citep{CalhounEtal14, warnick2018bayesian}. \cite{qiao2020doubly} recently proposed a model for sparse and irregularly sampled data for EEG, also based on prior evidence of dynamic changes in functional connectivity\citep{CabralEtAl14}. In this work we are concerned with static graphs. For instance, the EEG experiment we consider has often been analyzed with the goal of identifying the constant, persistent differences in EEG activation networks between alcoholic and control groups. Thus we follow the same goals here for more direct comparisons with both \cite{qiao2019functional} and \cite{zhu2016bayesian}. Further, in our study of the effect of brain injury on structural connectivity, we are primarily interested in this structure in the chronic phase after injury, in which case static, cross-sectional, population-level differences between the TBI and control groups are more clinically meaningful since they represent long-term changes that are more easily summarized. 

\hspace*{12pt} In this paper, we propose two different regularization schemes for functional graphical models. The first approach we consider is a direct Bayesian version of the frequentist functional graphical lasso\citep{qiao2019functional}. We propose also a functional graphical horseshoe, due to the horseshoe's known improvements upon the lasso's tendency to over-shrink large coefficients and under-shrink small coefficients in high-dimensional problems \citep{wang2012bayesian, li2019graphical}. Whereas most existing Bayesian approaches to covariance or precision matrix estimation assume structure such as banded covariance\citep{BanerjeeGhosal14} or decomposable graphs\citep{RajarEtAl08, XiangEtAl15, zhu2016bayesian}, neither the Bayesian functional graphical lasso nor the functional graphical horseshoe assume any structure other than sparsity. We provide Gibbs sampling algorithms for both of our proposed models, exploiting auxiliary variables to produce a set of easily-sampled conditional distributions. Through extensive simulation studies, we evaluate both the classification accuracy and fidelity of the estimated coefficients. We apply our proposed Bayesian functional graphical horseshoe to two motivating datasets, an EEG alcoholic versus control study\citep{qiao2019functional} and a novel study of white matter connectivity between healthy patients and those with a history of traumatic brain injury using data obtained from the Alzheimer's Disease Neuroimaging Initiative (ADNI).

\hspace*{12pt} This article is organized as follows: In Section \ref{sec:background}, we provide a brief background for Gaussian graphical models, including basic concepts and notations for structure learning through precision matrix estimation as well as the original graphical lasso \citep{yuan2007model}. The Section also reviews functional principal component analysis (FPCA) and its connection to both the frequentist and Bayesian regularization approaches to functional graphical models. In Section \ref{sec:prop}, we present both of our proposed approaches, the Bayesian functional graphical lasso and the functional graphical horseshoe model, as well as the easily-implemented Gibbs samplers. We present and discuss statistical and computational performance of our proposed methods in Section \ref{sec:sims}, followed by applying the functional graphical horseshoe to the motivating applications in Section \ref{sec:apps}. A summary and concluding remarks may be found in Section \ref{sec:summary}. The code for implementing our proposed algorithms is available at \url{https://github.com/jjniu/BayesFGM}.

\section{Background}\label{sec:background}

\subsection{Gaussian Graphical Models and the Graphical Lasso}\label{sec:GGM}
Suppose the random vector $\B{y} = (y_1, \cdots, y_p)^T$ follows a multivariate Gaussian distribution with mean $\B{\mu}$ and covariance matrix $\B{\Sigma}$. Then we define $\B{\Theta} = \B{\Sigma}^{-1}$ as the precision matrix, or concentration matrix. A Gaussian graphical model (GGM) is based on an undirected graph $G = (V,E)$, where $V = \{1,\dots, p\}$ is a non-empty set of vertices and $E\subseteq \{(i,j), i<j\}$ is a set of edges representing unordered pairs of vertices (also called nodes). Each variable $y_i$ is represented by a node in the graph, and $E$ determines the precision matrix in that, for $i \neq j$, $(\B{\Theta})_{ij} \neq 0$ if and only if $(i, j) \in E$. We thus have that $E$ encodes a Markov property in the distribution\citep{RueHeld05}. Letting $\mathcal{N}(i) = \{j: (i, j) \in E\}$ and adopting the convention that $\B{y}_{\mathcal{A}} = (y_j : j \in \mathcal{A})^T$ for a set of indices $\mathcal{A}$, we have that, for any node $i$, $y_i | \B{y}_{(-i)} \stackrel{d}{=} y_i | \B{y}_{\mathcal{N}(i)}$. Extending this with $V = \{1, \ldots, p\}$, it follows that $y_u \indep y_v | \mathbf{y}_{V\backslash\{u,v\}}$ if and only if $(\B{\Theta})_{uv} = 0$, where $\indep$ denotes statistical independence. This is the {pairwise Markov property}. By this property, learning the graph associated with a Gaussian graphical model is equivalent to estimating the precision matrix of the multivariate Gaussian distribution, making it a covariance estimation problem \citep{Dempster72}.

Given a sample $\B{y}_i, ~i= 1, \ldots, n$, stored in a data matrix $\B{Y} = (\B{y}_1 \cdots \B{y}_n)^T$, the goal is to estimate and select non-zero elements of $\B{\Theta}$, thereby obtaining an estimate of the undirected graph associated with the GGM. The log-likelihood of $\B{\Theta}$ (up to an additive constant) can be written as 
\begin{equation}
    \label{eq: GMM_ll}
    l(\mathbf{\Theta}) = \log \det \mathbf{\Theta} -\tr\left(\mathbf{S\Theta}/n\right),
\end{equation}
where $\mathbf{S} = \mathbf{Y}^T\mathbf{Y}$. The quantity $-l(\mathbf{\Theta})$ is a convex function of $\mathbf{\Theta}$ and the maximum likelihood estimator of $\mathbf{\Sigma}$ is $\widehat{\mathbf{\Sigma}} = \mathbf{S}/n$. This estimator enjoys nice properties such as consistency, but can be unstable when $p \approx n$. Further, even when $\widehat{\mathbf{\Sigma}}^{-1}$ exists, it can be an unsatisfactory estimator of $\B{\Theta}$ due to the fact that it will generally not be sparse, even if $\widehat{\mathbf{\Sigma}}$ is sparse. 

To find a more stable estimator of $\B{\Theta}$ that is simultaneously sparse, \cite{yuan2007model} proposed to solve a lasso-type regularized version of the likelihood objective function by finding 
\begin{equation}
\label{eq:objective_lasso}
\widehat{\mathbf{\Theta}} = \argmin_{\mathbf{\Theta}\in M^+} ~~\left\{-\log \det \mathbf{\Theta} + \tr\left(\mathbf{S\Theta}/n\right) + \lambda\|\mathbf{\Theta}\|_1\right\}, 
\end{equation}
where $M^+$ is the space of $p\times p$ symmetric positive definite matrices, the norm $\|\cdot\|_1$ is the sum of the absolute values of the off-diagonal elements, and $\lambda$ is a non-negative tuning parameter to control the number of zeros in the estimated precision matrix. This is a semi-definite programming problem for the precision matrix $\bm{\Theta}$.
\cite{yuan2007model} solved this problem with the so-called maxdet algorithm \citep{VanEtAl98}, 
while \cite{friedman2008sparse} proposed a more efficient coordinate descent algorithm for solving \eqref{eq:objective_lasso}. This is the {graphical lasso}, an approach that has since become very popular for structure learning in GGMs. 

\cite{wang2012bayesian} considered the fully Bayesian version of the graphical lasso by recognizing that solving \eqref{eq:objective_lasso} is equivalent to finding the maximum a posteriori (MAP) estimator in the following model,
\begin{eqnarray}\label{eq:bayesGlasso}
p(\B{y}_i \mid \B{\Theta}) &=& N(\B{y}_i \mid \B{0}, ~\B{\Theta}^{-1}), ~~i= 1, \ldots, n\\
p(\B{\Theta} \mid \lambda) &\propto& \prod_{i < j}DE(\theta_{ij} \mid \lambda) \prod_{i=1}^p Exp\left(\theta_{ii} \mid \frac{\lambda}{2}\right), ~~\B{\Theta} \in M^+,
\end{eqnarray}
where $N(\cdot | \B{0}, \B{\Theta}^{-1})$ denotes the density of a $N(\B{0}, \B{\Theta}^{-1})$ distribution, and likewise for the double exponential ($DE$) and exponential ($Exp$) distributions. Using a hierarchical representation of this model \citep{kyung2010penalized} and matrix partitioning techniques\citep{BanerjeeEtAl08, friedman2008sparse}, \cite{wang2012bayesian} developed an efficient Gibbs sampler for exploring the full posterior distribution and thus was able to extensively compare the results of the MAP and posterior mean estimators. 

\subsection{Functional Principal Component Analysis}\label{sec: FPCA}

For subject $i, ~i= 1, \ldots, n$, let the underlying, infinite dimensional function of interest be denoted $g_i(t), ~t \in \mathcal{T}$. We assume that $g_1, \ldots, g_n$ are identically distributed and independent zero-mean functions in $L^2[\mathcal{T}]$ with covariance function $\cov(g_j(s), g_j(t)) =: \Sigma(s,t), ~(s, t) \in \mathcal{T} \times \mathcal{T}$, where $\mathcal{T}$ is a compact interval on the real line. \cite{Kar46} and \cite{Loe46} independently discovered the functional principal component analysis (FPCA) expansion \citep{bosq2012linear},
\begin{equation}
    \label{eq:KL_expansion}
    g_{i}(t) = \sum_{k = 1}^{\infty} a_{ik}\phi_{k}(t),
\end{equation}
where $\{\phi_{k}(t)\}_{k = 1}^{\infty}$ are the orthonormal set of eigenfunctions with corresponding eigenvalues $\{\lambda_{k}\}_{k=1}^\infty$ satisfying  $\Sigma(s,t) =\sum_{k=1}^{\infty}\lambda_{k}\phi_{k}(s)\phi_{k}(t)$, by Mercer's Theorem, and $a_{ik} = \int g_{i}(t)\phi_{k}dt$ are the functional principal component (FPC) scores of $g_i$, uncorrelated across $k$ with $E(a_{ik}) = 0$ and $\var(a_{ik}) = \lambda_k$. By assumption, the $a_{ik}$ are independent across $i$. Like ordinary principal components analysis \citep{Jolliffe02}, the expansion can be truncated to obtain a finite-dimensional approximation to the infinite-dimensional process. In what follows, the proposed functional graphical models can work with any basis expansion (e.g., wavelets or Fourier), but we use FPCA due to the mean square optimality of the truncated approximation.

Performing FPCA in practice amounts to finding the spectral decomposition of an approximation to the covariance function. When $g_i, ~i= 1, \ldots, n$, are observed on the same evenly spaced grid $t_1, \ldots, t_m$ independent of $i$, this amounts to standard singular value decomposition of the sample covariance matrix. For irregularly spaced functions and/or different numbers of observations on each function, SVD will likely provide a poor approximation to the true eigensystem associated with $\Sigma(\cdot, \cdot)$. In this case, the PACE algorithm\citep{yao2005functional} can be used for performing FPCA via conditional expectation. In our applications, we use SVD for the EEG example and PACE for the diffusion MRI example, as the latter involves irregularly sampled longitudinal data.

\subsection{Functional Graphical Models}\label{sec:FGM}
For a particular subject $i$, suppose we (discretely) observe $p$ functions $g_{i1}(t), \ldots, g_{ip}(t)$ where $g_{ij}$ is the subject's function observed on node $j$. Suppose further that each function is a Gaussian process so that $(g_{i1}, \ldots, g_{ip})$ is a realization from a $p-$dimensional multivariate Gaussian process (MGP). As in typical GGMs, we associate to the MGP an undirected graph $G = (V, E)$ that represents the conditional dependence network. Here, conditional dependence of the functions $g_{ij}$ and $g_{ij^\prime}$ is in terms of the cross-covariance function,
\begin{equation}
    C_{jj^\prime}(s,t) = \cov\left(g_{ij}(s),g_{ij^\prime}(t)~|~{g_k(\cdot), k\neq j,j^\prime}\right),
\end{equation}
assumed to be the same for $i= 1, \ldots, n$. 

With the covariance function in hand, we can use FPCA and approximate each $g_{ij}$ with the $M$-dimensional truncation,
\begin{equation}
\label{eq:m-truncated}
g_{ij}^M(t) = \sum_{k = 1}^{M} a_{ijk}\phi_{jk}(t), ~~M < \infty.
\end{equation}
The function for subject $i$ at node $j$ can thus be represented with the coefficient vector $\bm{a}^M_{ij} = (a_{ij1},\dots, a_{ijM})^T$, so that each subject's entire functional information over all $p$ nodes is encoded in  $\bm{a}^M_{i} = ((\bm{a}_{i1}^M)^T, \dots, (\bm{a}_{ip}^M)^T)^T \in \mathbb{R}^{Mp}$. Under the Gaussian assumption and independently observed subjects, the Kahrunen-Lo\'{e}ve Theorem  tells us that $\bm{a}^M_{i} \stackrel{\text{iid}}{\sim} N_{Mp}(\B{0}, (\B{\Theta}^M)^{-1})$. For learning the graphical model, \cite{qiao2019functional} show that, in the finite-rank case (in which the $M$-truncated approximation is exact),
\begin{equation}
\label{eq:fedge}
    {E}^M = \left\{ (i,j): \|\bm{\Theta}_{ij}^M\|_F\neq 0, (i,j)\in V^2, i\neq j \right\},
\end{equation}
where $\bm{\Theta}_{ij}^M$ is the $M \times M$ block submatrix of $\B{\Theta}^M$ corresponding to the node pair $(i, j) \in V \times V$ and $\|\cdot\|_F$ is the Frobenius norm. Thus, structure learning in the functional graphical model is equivalent to finding the $(i, j)$ pairs for which $\|\bm{\Theta}_{ij}^M\|_F\neq 0$.

The connection in \eqref{eq:fedge} to the graphical lasso and the group lasso \citep{yuan2006model} led to estimating the graph from functional data with\citep{qiao2019functional} 
\begin{equation}
\label{eq: fglasso}
 \hat{\bm{\Theta}}^M = \argmax_{\bm{\Theta}^M} \left\{\log{\det{\bm{\Theta}^M} }-\tr\left(\bm{S}^M \bm{\Theta}^M\right)- \lambda_n \sum_{i\neq j} \|\bm{\Theta}_{ij}^M\|_F \right\}
\end{equation}
where $\B{S}^M$ is the sample covariance matrix computed from estimated FPC scores $\widehat{\B{a}}_i^M \in \mathbb{R}^{Mp}$, found via SVD or otherwise, and $\lambda_n > 0$ is a tuning parameter. As with the group lasso, blockwise sparsity is achieved as $\lambda_n \rightarrow \infty$. This is the {functional graphical lasso} (fglasso)\citep{qiao2019functional} . The edge set of the estimated graph is then $\hat{E}^M = \left\{ (i,j): \|\hat{\bm{\Theta}}_{ij}^M\|_F\neq 0, (i,j)\in V^2, i\neq j \right\}.$ The superscript $M$ here (and hereafter) is used to denote the parameter is dependent on $M$. But for simplicity, we may omit the superscripts where the context is clear. Rather than using identical truncated number $M$ across $j = 1, \dots, p$, one can select $M_j$ separate for each $j$, as different functional variables may have different smoothness levels. \cite{qiao2019functional} show that the fglasso enjoys model selection consistency, and provide a block coordinate descent algorithm for optimizing the objective function.

\section{Bayesian Functional Graphical Models}\label{sec:prop}
\subsection{The Bayesian Fglasso}\label{sec:BayesFglassoMod}
It is well known that frequentist optimization of objective functions may often be viewed as maximum a posteriori (MAP) estimation under a Bayesian model, provided there exists a prior density corresponding to the penalty term in the objective function. For the fglasso objective function in \eqref{eq: fglasso}, the Bayesian counterpart uses a prior on the precision matrix given by
\begin{equation}\label{eqn:groupPrior}
    \pi(\mathbf{\Theta})\propto \exp\left\{-\lambda\sum_{i\neq j}\|\mathbf{\Theta}_{ij}\|_{F}\right\},
\end{equation}
where $\|\cdot\|_F$ denotes the Frobenius norm and $\mathbf{\Theta}_{ij}\in\mathbb{R}^{M\times M}$ is the $(i,j)$th submatrix in $\mathbf{\Theta}$ associated with the conditional cross-correlation between node $i$ and node $j$, $i, j= 1, \ldots, p; ~i\neq j$.  Since the precision matrix is symmetric, we need only to consider the upper off-diagonal elements for computational simplicity. As used in the Bayesian group lasso hierarchical representation \citep{kyung2010penalized}, we have the following identity,
 \begin{equation}
     \begin{aligned}
        \exp(-\lambda\|\mathbf{\Theta}_{ij}\|_F) = \int_0^{\infty}\left(\frac{1}{2\pi\tau_{ij}^2}\right)^{\frac{M^2}{2}}\exp\left(-\frac{\|\mathbf{\Theta}_{ij}\|^2_F}{2\tau_{ij}^2}\right)
        \frac{\left(\frac{\lambda^2}{2}\right)^{\frac{M^2+1}{2}}(\tau_{ij}^2)^{\frac{M^2+1}{2}-1}}{\Gamma\left(\frac{M^2+1}{2}\right)} \exp\left(-\frac{\lambda^2\tau_{ij}^2}{2}\right)d\tau_{ij}^2.\\
    \end{aligned}    
\end{equation} 
Thus, we can rewrite $\pi(\mathbf{\Theta})$ as a scale mixture of a multivariate normal distribution on the off-diagonal elements. Without loss of generality, suppose $M$ does not depend on node $j$, and let $\mathbf{\omega}_{ij} = \text{vec}(\mathbf{\Theta}_{ij}) \in \mathbb{R}^{M^2}$, for $i, j = 1, \ldots, p, ~i \neq j$. Then we can introduce the auxiliary latent parameters $\mathbf{\tau} = (\tau_{ij})$, so the prior in \eqref{eqn:groupPrior} can be attained as a gamma mixture of normals, leading to the functional graphical lasso hierarchy
\begin{equation}\label{eqn:fgm_hier}
    \mathbf{\omega}_{ij}|\tau_{ij}^2\sim N_{M^{2}}(\mathbf{0}, \tau_{ij}^2\mathbf{I}_{M^2});  ~~\tau_{ij}^2 \sim \mbox{Gamma}\left(\frac{M^2+1}{2}, \frac{\lambda^2}{2}\right).
\end{equation}

We assume the basis expansion is a lossless or approximately lossless representation from the raw data $g_{ij}(t)$ to $\mathbf{a}_{ij}^M$, where isomorphic transformation ensures that any basis coefficients can be considered as transformed raw data rather than estimated parameters \citep{morris2011automated}. Denote by $\hat{\mathbf{a}}_{i} = (\hat{\mathbf{a}}_{i1}^T, \ldots, \hat{\mathbf{a}}_{ip}^T)^T \in \mathbb{R}^{Mp}$ the estimated $M$-truncated functional principal component scores for the observed functions on sample $i$, $g_{i1}(\cdot), \ldots, g_{ip}(\cdot)$. When $(g_{i1}(\cdot), \ldots, g_{ip}(\cdot))$ are drawn from an MGP, $\hat{\mathbf{a}}_{i}$ follows an $Mp$-dimensional Gaussian distribution. Then the Bayesian fglasso model can be expressed as
\begin{equation}
    \label{eq:fglasso_bayesian_eq}
    \begin{aligned}
    &p(\hat{\mathbf{a}}_{i}|\mathbf{\Theta}) = N_{Mp}(\hat{\mathbf{a}}_{i}|\boldsymbol{0}, {\mathbf{\Theta}}^{-1}), \quad i = 1,\dots, N\\
    &p(\mathbf{\Theta}|\lambda) = \frac{1}{C} \prod_{\ell=1}^{Mp}Exp\left(\theta_{\ell \ell} | \frac{\lambda^2}{2}\right)
    \prod_{i<j} N_{M^{2}}(\mathbf{\omega}_{ij}|\mathbf{0}, \tau_{ij}^2\mathbf{I}_{M^2})\mbox{Gamma}\left(\tau_{ij}^2|\frac{M^2+1}{2},\frac{\lambda^2}{2}\right),\\
    \end{aligned}
\end{equation}
where $\theta_{11}, \ldots, \theta_{pp}$ are the diagonal elements of $\mathbf{\Theta}$ and $C$ is a normalizing constant.

The hierarchical representation in \eqref{eqn:fgm_hier} facilitates the use of conditional conjugacy in deriving a block Gibbs sampler for exploring the posterior distribution. For a fixed regularization parameter $\lambda$, the posterior distribution associated with the Bayesian fglasso model \eqref{eq:fglasso_bayesian_eq} is given by
 \begin{equation}
     \begin{aligned}
          p(\mathbf{\Theta}, \mathbf{\tau}^2|\mathbf{S}, \lambda)\propto &|\mathbf{\Theta}|^{\frac{n}{2}}\exp\left\{-\tr(\frac{1}{2}\mathbf{S\Theta})\right\}\prod_{\ell=1}^{Mp}\frac{\lambda^2}{2}\exp\left(-\frac{\lambda^2}{2}\theta_{\ell \ell}\right)\\
          &\times \prod_{i<j}\left(\frac{1}{2\pi\tau_{ij}^2}\right)^{\frac{M^2}{2}}\exp\left(-\frac{\|\mathbf{\Theta}_{ij}\|^2_F}{2\tau_{ij}^2}\right)\\
          &\times \left\{\frac{\left(\frac{\lambda^2}{2}\right)^{\frac{M^2+1}{2}}(\tau_{ij}^2)^{\frac{M^2+1}{2}-1}}{\Gamma\left(\frac{M^2+1}{2}\right)}\exp\left(-\frac{\lambda^2\tau_{ij}^2}{2}\right)\right\}1_{\mathbf{\Theta}\in M^+},\\  
     \end{aligned}
 \end{equation}
where $\mathbf{S} = \sum_{i=1}^n \mathbf{\hat{a}}_i \mathbf{\hat{a}}_i^T$ is the sample scatter matrix of the functional PC scores. This representation allows us to adapt the block Gibbs sampling scheme proposed by \cite{wang2012bayesian}.
 
By assumption, the off-diagonal entries of each block submatrix on the main diagonal, $\mathbf{\Theta}_{ii}, ~i= 1, \ldots, p$, are all zeros. Partition the precision and sample fpc score covariance matrix as follows: 
\begin{equation}\label{eqn:ThetaPart}
         \mathbf{\Theta} = \begin{bmatrix}
    \mathbf{\Theta}_{11}& \mathbf{\theta}_{12} \\ 
    \mathbf{\theta}_{12}^T& \theta_{22} 
    \end{bmatrix}, ~
    \mathbf{S} = \begin{bmatrix}
        \mathbf{S}_{11}& \mathbf{s}_{12} \\ 
        \mathbf{s}_{12}^T& s_{22} 
    \end{bmatrix}.
\end{equation}
We define 
\begin{equation}
    \mathbf{\theta}_{12} = \begin{bmatrix}
        \bar{\mathbf{\theta}}_{12} \\ 
         \mathbf{0}
    \end{bmatrix},\\
\end{equation}
with $\mathbf{\Theta}$ permuted so that the last column / row corresponds to node $j$ and score $k$, and $\bar{\mathbf{\theta}}_{12} = Cov(\hat{a}_{jk}, (\hat{\mathbf{a}}_1, \ldots, \hat{\mathbf{a}}_{j-1}, \hat{\mathbf{a}}_{j+1}, \ldots, \hat{\mathbf{a}}_p)^T) \in \mathbb{R}^{M(p-1)}$ with $\hat{\mathbf{a}}_{k} \in \mathbb{R}^M$ the collection of fpc scores at node $k$. The $\mathbf{0} \in \mathbb{R}^{M-1}$ vector follows from $\hat{a}_{jk}$ being uncorrelated with other scores at node $j$.

Define $\mathbf{T} = \left(\tau^2_{ij}\right)_{p\times p}\otimes \mathbf{J}_{M\times M}$ where $\tau_{ii} = 0$ for $i= 1, \ldots, p$ and $\mathbf{J}_{M\times M} = \mathbf{11}^T$ is the matrix with all ones. We similarly partition it as
\begin{equation}\label{eqn:TPart}
    \mathbf{T} = \begin{bmatrix}
        \mathbf{T}_{11}& \mathbf{t}_{12} \\ 
        \mathbf{t}_{12}^T& 0
    \end{bmatrix},
\end{equation}
where 
\begin{equation}
    \mathbf{t}_{12} = \begin{bmatrix}
    \bar{\mathbf{t}}_{12} \\ 
     \mathbf{0}
    \end{bmatrix}\\
\end{equation}
with $\bar{\mathbf{t}}_{12} \in \mathbb{R}^{M(p-1)}$ defined analogously to $\bar{\mathbf{\theta}}_{12}$.

The conditional distribution of the nonzero variables in the last column (or row) of $\mathbf{\Theta}$ is
\begin{equation}
    \begin{aligned}
    p(\bar{\mathbf{\theta}}_{12}, \theta_{22}&|\mathbf{\Theta}_{11}, \mathbf{T}, \mathbf{S}, \lambda) \propto (\theta_{22}-\bar{\mathbf{\theta}}_{12}^T\overline{\mathbf{\Theta}_{11}^{-1}}\bar{\mathbf{\theta}}_{12})^{\frac{n}{2}}\\
    &\times\exp\left\{-\frac{1}{2}[\bar{\mathbf{\theta}}_{12}^T\mathbf{D}_{\tau}^{-1}\bar{\mathbf{\theta}}_{12}+2\bar{\mathbf{s}}_{12}^T\bar{\mathbf{\theta}}_{12}+(s_{22}+\lambda^2)\theta_{22}]\right\}\\
    \end{aligned}
\end{equation}
where $\mathbf{D}_{\tau} = \mbox{diag}(\bar{\mathbf{t}}_{12})$ and $\overline{\mathbf{\Theta}_{11}^{-1}} \in \mathbb{R}^{M(p-1) \times M(p-1)}$ is the cross covariance matrix associated with the remaining $p-1$ nodes. We make a change of variables,
$\mathbf{\beta} = \bar{\mathbf{\theta}}_{12}, \gamma = (\theta_{22}-\bar{\mathbf{\theta}}_{12}^T\overline{\mathbf{\Theta}_{11}^{-1}}\bar{\mathbf{\theta}}_{12})$, and denote $\mathbf{C} = (\mathbf{D}_{\tau}^{-1}+(s_{22}+\lambda^2)\overline{\mathbf{\Theta}_{11}^{-1}})^{-1}$. 
This implies
\begin{equation}
    \label{eq:beta_gamma}
    \mathbf{\beta}, \gamma|\mathbf{\Theta}_{11}, \mathbf{T}, \mathbf{S},\lambda \sim N_{M(p-1)}(-\mathbf{C}\bar{\mathbf{s}}_{21}, \mathbf{C})\mbox{Gamma}\left(\frac{n}{2}+1, \frac{s_{22}+\lambda^2}{2}\right).
\end{equation}
All elements in the matrix $\mathbf{\Theta}$ can be sampled by sampling one row and column at a time, permuting $\mathbf{\Theta}$ after each iteration. Due to the structure of $\hat{\mathbf{a}}_{i}$, we first cycle through all columns corresponding to the same node, then move to next node. 

After complete updating of all the off-diagonal elements, the diagonal elements of $\mathbf{\Theta}$ and the shrinkage parameters $\tau_{ij}$ need to be sampled. The full conditional  distributions of $(\tau_{ij}^2)^{-1}$ are seen to be independently inverse Gaussian with mean $\sqrt{\frac{\lambda^2}{\|\mathbf{\Theta}_{ij}\|^2_F}}$ and shape $\lambda^2$. Put another way, the reparameterized model based on one particular permutation of $\mathbf{\Theta}$ under the Bayesian functional graphical lasso is 
\begin{equation}
\label{eq: fglasso_summary}
\begin{aligned}
    \mathbf{\beta}&|\mathbf{\Theta}_{11}, \mathbf{T}, \mathbf{S},\lambda\sim N_{M(p-1)}(-\mathbf{C}\bar{\mathbf{s}}_{21}, \mathbf{C})\\
    \gamma&|\mathbf{S}, \lambda\sim \mbox{Gamma}\left(\frac{n}{2}+1, \frac{s_{22}+\lambda^2}{2}\right) \\
    \frac{1}{\tau_{ij}^2}&|\mathbf{\Theta}_{ij},\lambda
\stackrel{indep.}{\sim} \mbox{Inverse Gaussian}\left(\sqrt{\frac{\lambda^2}{\|\mathbf{\Theta}_{ij}\|^2_F}}, \lambda^2\right), ~~i, j= 1, \ldots, p; ~i \neq j.\\
\end{aligned}
\end{equation}
Since $\gamma>0$ with probability one, the positive definite constraint on $\mathbf{\Theta}$ is maintained in each iteration. The argument for the functional case is adapted from that given by \cite{wang2012bayesian}. Suppose at the current iteration the sample $\mathbf{\Theta}^{(c)}$ is positive definite, so all its $pM$ corresponding leading principal minors are positive. After updating the particular column and row of $\mathbf{\Theta}$ by sampling $\mathbf{\beta}$ and $\gamma$ by \eqref{eq: fglasso_summary}, the new sample $\mathbf{\Theta}^{(c+1)}$ has the same leading principal minors as $\mathbf{\Theta}^{(c)}$ except the one corresponding to the updated column/row, which is of order $pM$. It is easy to find that this last leading principal minor is $\det(\mathbf{\Theta}^{(c+1)}) = \gamma\det(\mathbf{\Theta}^{(c)}_{11})$, where $\det(\mathbf{\Theta}^{(c)}_{11})$ is the $(pM-1)^{th}$ leading principal minor of $\mathbf{\Theta}^{(c)}$ excluding the updated column and row. Thus $\gamma > 0$ means that $\det(\mathbf{\Theta}^{(c+1)}) > 0$ and all leading principal minors of the updated matrix are positive. To ensure each MCMC realization $\mathbf{\Theta}^{(m)} \in M^+$ for $m = 0, 1, 2, \ldots$, it is only required that the chain is initialized with $\mathbf{\Theta}^{(0)} \in M^+$. Algorithm \ref{algorithm:Bayesfglasso_Gibbs} details the Bayesian fglasso Gibbs sampler.

\begin{algorithm}[tb]
\SetAlgoLined
\textbf{Input:}  Sum of the products matrix $\bm{S}$, i.e., $\bm{S} = \bm{Y}^T\bm{Y}$.\\
\textbf{Output:} MCMC sample of the precision matrix $\bm{\Theta}^{(1)}, \ldots, \bm{\Theta}^{(L)}$.\\ 
\textbf{Initialization}: Set $p$ to be number of nodes in graph, set initial values $\bm{\Theta}^{(0)} = \bm{I}$, $\bm{\Sigma}^{(0)} = \bm{I}$, $\bm{T}^{(0)} = \bm{J}$, where $\bm{I}$ is $pM\times pM$ identity matrix and $\bm{J}$ is a $pM\times pM$ matrix with all elements equal to one\;
 \While{Convergence criteria are not met}{
 \For{$i = 1,\dots, p$}{
        Partition $\bm{\Theta}^{(l)}$, $\bm{S}$ and $\bm{T}^{(l)}$ into $p\times p$ blocks (focus on updating $i$th column block of $\bm{\Theta}$ corresponding node $i$)\;
        \For{$j = 1,\dots, M$}{
            1. Partition $\bm{\Theta}^{(l)}$, $\bm{S}$ and $\bm{T}^{(l)}$ as in \eqref{eqn:ThetaPart} and \eqref{eqn:TPart}\;
            2. Draw $\gamma^{(l+1)} \sim \mbox{Gamma}\left(\frac{n}{2}+1, \frac{s_{22}+\lambda}{2}\right)$\;
            3. Draw $\bm{\beta}^{(l+1)} \sim N_{(p-1)M}(-\bm{C}^{(l)}\bar{\bm{s}}_{21}, \bm{C}^{(l)})$,where $\bm{C}^{(l)} = ((\bm{D}_{\tau}^{(l)})^{-1}+(s_{22}+\lambda)\overline{\bm{\Theta}_{11}^{-1}}^{(l)})^{-1}$\;
            4. Update $\bm{\theta}_{21}^{(l+1)} = (\bm{\beta}^{(l+1)},\bm{0}),\bm{\theta}_{12}^{(l+1)} = (\bm{\theta}_{21}^{(l+1)})^T, \theta_{22}^{(l+1)} = \gamma^{(l+1)}+(\bm{\beta}^{(l+1)})^T\overline{\bm{\Theta}_{11}^{-1}}^{(l+1)}\bm{\beta}^{(l+1)}$\;
        }
 Update $\bm{T}^{(l+1)}$ by sampling $(1/\tau_{ij}^2)^{(l+1)}|\bm{\Theta}^{(l+1)},\lambda\sim \mbox{Inverse Gaussian}\left(\sqrt{\frac{\lambda^2}{\|\bm{\Theta}_{ij}^{(l+1)}\|^2_F}}, \lambda^2\right)$ for $i,j = 1,\dots,p$
 } 
 Store the realization of precision matrix $\bm{\Theta}^{(l+1)}$\;
 Increment $l \leftarrow l+1$.
 }
\caption{Bayesian functional graphical lasso Gibbs sampler}
\label{algorithm:Bayesfglasso_Gibbs}
\end{algorithm}

Given the MCMC output of a sample of precision matrices, $\mathbf{\Theta}^{(1)}, \ldots, \mathbf{\Theta}^{(L)}$, several inferential procedures are possible for constructing an estimate of $\mathbf{\Theta}$. Continuous shrinkage priors do not put positive probability mass on exact zeros in the precision matrix, and is has been argued that using (non zero) posterior means as the basis for inference is often preferable to binary thresholding due to the estimator's optimality under squared error loss\citep{carvalho2010horseshoe}. Nevertheless, it is sometimes necessary to produce a sparse estimate with exact zeros, especially in the case of graphical models. \cite{carvalho2010horseshoe} and \cite{wang2012bayesian} discuss some possible thresholding rules. In our case, we construct the precision matrix (and thus graph) estimate by Bayesian false discovery rate (FDR) based inference or confidence regions of $\{\hat{\bm{\Theta}}_{ij}\}$, which is discussed in Section \ref{sec:fdr}.

The Bayesian fglasso proposed here assumes that the regularization parameter $\lambda$ is fixed, meaning that it must be tuned and selected {a priori}. Cross-validation is computationally expensive, especially for Bayesian models implemented via MCMC. Further, it has been shown that cross-validation based on the log-likelihood loss function tends to lead to overfitting and unnecessarily dense graphs\citep{wasserman2009high}. Other than cross validation, approaches such as Akaike information criterion (AIC), Bayesian information criterion (BIC),  and stability selection \citep{meinshausen2006high} have been well studied in the graphical model literature. In the functional case, though, AIC/BIC does not work well, since it is unclear how to calculate the effective degrees of freedom. Thus, selecting an appropriate hyperparameter $\lambda$ ahead of time is a nontrivial task. On the other hand, in the Bayesian framework, we can (for instance) assign a gamma prior $\lambda^2\sim \mbox{Gamma}(s,r)$. In this case, the full conditional for $\lambda$ is
\begin{equation}\label{eq:lasso_lambda}
    \lambda^2|\mathbf{T}, \mathbf{\Theta}\sim \mbox{Gamma}\left( 
s + pM + \frac{p(p-1)(M^2+1)}{4}, ~r+\frac{\sum_{l}\theta_{ll}+\sum_{i<j}\tau_{ij}^2}{2}\right).
\end{equation}
This can in turn be incorporated into the Gibbs sampler given in Algorithm \ref{algorithm:Bayesfglasso_Gibbs} as an additional sampling step. 

In general, different functional variables may have different smoothness levels, in which case we can take different $M_j$ across $j = 1, \dots, p$. With different truncation levels, $\mathbf{\Theta}_{ij}$ is a rectangle block with size $M_i
\times M_j$ and $\mathbf{\Theta}$ has dimension $\sum_{j = 1}^p M_j$. It is straightforward to modify the algorithm with nonsquare blocks. The full conditional for $\lambda$ is updated as
\begin{equation}\label{eq:lasso_lambda_diff_M}
    \lambda^2|\mathbf{T}, \mathbf{\Theta}\sim \mbox{Gamma}\left( 
s + \sum_j M_j + \sum_{i < j} \frac{M_iM_j - 1}{2}, ~r+\frac{\sum_{l}\theta_{ll}+\sum_{i<j}\tau_{ij}^2}{2}\right).
\end{equation}

\subsection{The Functional Graphical Horseshoe}\label{sec:horse}
In the presence of sparsity, as is often the case for precision matrices associated with GGMs, it is desirable to have a shrinkage approach that yields exact or values close to zero for the true null cases while simultaneously shrinking the truly non-zero cases as little as possible to minimize the resulting bias. To address this desire, \cite{carvalho2010horseshoe} proposed the horseshoe prior. The prior has high probability concentration near zero and and is heavy-tailed, properties that contribute to desired shrinkage behavior. Further, the prior can be expressed as a scale mixture of Gaussian distributions and thus is easily incorporated into a Gibbs sampler for posterior exploration. 
The horseshoe was originally proposed for the sparse normal means model. It was recently extended to estimation of GGMs\citep{li2019graphical}, where it was established that the horseshoe estimators are close to be unbiased least-square estimators with high probability and, further, that the Bayesian graphical lasso tends to be further away from the least squares estimates than the graphical horseshoe. In this section, we propose an extension of graphical horseshoe regularization to the case of functional graphical models.

We define the functional graphical horseshoe by using horseshoe priors on each off-diagonal block of the precision matrix and exponential priors on the diagonal elements. This yields the following prior:
\begin{equation}
    \begin{aligned}
        \theta_{\ell \ell} &\sim Exp(\lambda_{\ell \ell}^2/2), ~~\ell = 1, \ldots, pM\\
        \mathbf{\omega}_{ij}&\stackrel{indep.}{\sim} N_{M^2}(\mathbf{0},~ \lambda_{ij}^2\tau^2\mathbf{I}), ~~i, j= 1, \ldots, p, ~i \neq j\\
        \lambda_{ij}&\stackrel{iid}{\sim} C^+(0,1), ~~i, j= 1, \ldots, p; ~~~~\tau \sim C^+(0,1),
    \end{aligned}
\end{equation}
where $\mathbf{\omega}_{ij} = \text{vec}(\mathbf{\Theta}_{ij})$ and $C^+(0,1)$ represents the half-Cauchy distribution with density $p(x)\propto (1+x^2)^{-1}, ~x>0$. As in other versions of the horseshoe prior, the global shrinkage parameter $\tau$ is determined by the sparsity of the entire precision matrix, whereas the local shrinkage parameters $\lambda_{ij}$ preserves blocks with $\| \mathbf{\Theta}_{ij}\| \neq 0$ by allowing them to be pulled toward zero considerably less than the zero blocks. Unlike \cite{li2019graphical}, but similar to \cite{wang2012bayesian}, we specify an $Exp(\lambda_{\ell, \ell}^2/2)$ prior for the diagonal elements of $\mathbf{\Theta}$. This is convenient for deriving the full conditional distributions and does not affect inference since the graph is determined by the off-diagonal elements.

The full conditional distribution of $\mathbf{\Theta}$ under the assumption of multivariate Gaussian likelihood is given by
\begin{equation}
    \begin{aligned}
        p(\mathbf{\Theta}|\lambda, \Lambda, \tau, \mathbf{S})&\propto |\mathbf{\Theta}|^{\frac{n}{2}}\exp\left\{-\tr(\frac{1}{2}\mathbf{S\Theta})\right\}\prod_{l=1}\frac{\lambda_{ll}^2}{2}\exp\left(-\frac{\lambda_{ll}^2}{2}\theta_{ll}\right) \prod_{i<j}N_{M^2}(\mathbf{\omega}_{ij}|\mathbf{0}, \lambda_{ij}^2\tau^2\mathbf{I})C^+(\lambda_{ij}|0,1)1_{\mathbf{\Theta}\in M^+}.
    \end{aligned}
\end{equation}
where $\Lambda = \{\lambda_{ij}\}_{i,j=1}^p$. The standard technique for creating a straightforward Gibbs sampler with the horseshoe is to use the fact that if $x^2|a\sim \mbox{inverse Gamma}(1/2, 1/a)$ and $a\sim \mbox{inverse Gamma}(1/2, 1/A^2)$, then, marginally, $x\sim C^+(0,A)$ \citep{makalic2015simple}. Thus, we introduce latent variables $\nu_{ij}$ and $\zeta$ to facilitate conditional conjugacy when updating the shrinkage parameters $\lambda_{ij}$ and $\tau$. 

Under the parameter-expanded hierarchical model, the full conditional distribution of the precision matrix is given by
\begin{equation}
    \begin{aligned}
        p(\mathbf{\Theta}|\mathbf{S}, \mathbf{\Lambda}, \tau, \mathbf{V}, \zeta)&\propto |\mathbf{\Theta}|^{\frac{n}{2}}\exp\left\{-\tr(\frac{1}{2}\mathbf{S\Theta})\right\}\prod_{l=1}^{Mp}\frac{\lambda_{\ell \ell}^2}{2}\exp\left(-\frac{\lambda_{\ell \ell}^2}{2}\theta_{ll}\right)\\
        &\times \prod_{i<j}N_{M^2}(\mathbf{\omega}_{ij}|\mathbf{0}, \lambda_{ij}^2\tau^2\mathbf{I})\prod_{i<j}\nu_{ij}^{-\frac{1}{2}}\lambda_{ij}^{-3}\exp\left(-\frac{1}{\lambda_{ij}^2\nu_{ij}}\right)\nu_{ij}^{-\frac{3}{2}}\exp\left(-\frac{1}{\nu_{ij}}\right)\\
        &\times \zeta^{-\frac{1}{2}}\tau^{-3}\exp\left(-\frac{1}{\tau^2\zeta}\right)\zeta^{-\frac{3}{2}}\exp\left(-\frac{1}{\zeta}\right).
    \end{aligned}
\end{equation}
We can use a data-augmented Gibbs sampler with the same matrix permutation as used for the Bayesian fglasso proposed in Subsection \ref{sec:BayesFglassoMod}.

In each iteration, the rows and columns of the $Mp-$dimensional matrices $\mathbf{\Theta}$, $\mathbf{S}$, $\mathbf{\Lambda} =\{\lambda^2_{ij}\}_{p\times p}\otimes \mathbf{J}_{M\times M}$, and $\mathbf{V}= \{\nu^2_{ij}\}_{p\times p}\otimes \mathbf{J}_{M\times M}$ are partitioned the same way as in Subsection \ref{sec:BayesFglassoMod} to derive the full conditional distributions; i.e.,
\begin{equation}
    \label{eq:partion_horseshoe}
    \begin{aligned}
            \mathbf{\Theta} = \begin{bmatrix}
             \mathbf{\Theta}_{11}& \mathbf{\theta}_{12} \\ 
             \mathbf{\theta}_{12}^T& \theta_{22} 
            \end{bmatrix},&~~
        \mathbf{S} = \begin{bmatrix}
        \mathbf{S}_{11}& \mathbf{s}_{12} \\ 
         \mathbf{s}_{12}^T& s_{22} 
        \end{bmatrix},\\
        \mathbf{\Lambda}= \begin{bmatrix}
        \mathbf{\Lambda}_{11}& \mathbf{\lambda}_{12} \\ 
         \mathbf{\lambda}_{12}^T& \lambda_{22} 
        \end{bmatrix},&~~
        \mathbf{V}= \begin{bmatrix}
        \mathbf{V}_{11}& \mathbf{\nu}_{12} \\ 
         \mathbf{\nu}_{12}^T& \nu_{22} 
        \end{bmatrix},
    \end{aligned}
\end{equation}
where the blocks are arranged as before. The derivation of full conditionals for the last column $\mathbf{\theta}_{12}$ and $\theta_{22}$ is similar to the Bayesian fglasso by changing variables. The conditional distribution of nonzero variables of the last column in $\mathbf{\Theta}$ is
\begin{equation}
    \begin{aligned}
    p(\bar{\mathbf{\theta}}_{12}, \theta_{22}&|-) \propto \left(\theta_{22}-\bar{\mathbf{\theta}}_{12}^T\overline{\mathbf{\Theta}_{11}^{-1}}\bar{\mathbf{\theta}}_{12}\right)^{\frac{n}{2}}
  \exp\left\{-\frac{1}{2}[\bar{\mathbf{\theta}}_{12}^T\mathbf{D}_{\tau}^{-1}\bar{\mathbf{\theta}}_{12}+2\bar{\mathbf{s}}_{12}^T\bar{\mathbf{\theta}}_{12}+(s_{22}+\lambda_{22}^2)\theta_{22}]\right\},\\
    \end{aligned}
\end{equation}where $\mathbf{D}_{\tau} = \tau^2\mbox{diag}(\bar{\mathbf{\lambda}}_{12})$.
Making a change of variables by $\mathbf{\beta} = \bar{\mathbf{\theta}}_{12}, \gamma = (\theta_{22}-\bar{\mathbf{\theta}}_{12}^T\overline{\mathbf{\Theta}_{11}^{-1}}\bar{\mathbf{\theta}}_{12})$, and letting $\mathbf{C} = (\mathbf{D}_{\tau}^{-1}+(s_{22}+\lambda_{22}^2)\overline{\mathbf{\Theta}_{11}^{-1}})^{-1}$ , the full conditional of $\mathbf{\beta}, \gamma$ is
\begin{equation}\label{eq: horse_beta_gamma}
    \mathbf{\beta}, \gamma|- \sim N_{(p-1)M}(-\mathbf{C}\bar{\mathbf{s}}_{21}, \mathbf{C})\mbox{Gamma}\left(\frac{n}{2}+1, \frac{s_{22}+\lambda_{22}^2}{2}\right)
\end{equation}
We first cycle through all columns corresponding to the same node, then move to next node.
After the entire $\mathbf{\Theta}$ is updated, the local and global shrinkage parameters $\lambda_{ij}$ and $\tau$ need to be sampled. Through conditional conjugacy, the full conditional distributions of $\lambda_{ij}, \nu_{ij}, \tau^2,$ and $\zeta$ are quickly seen to be inverse Gamma. The condition $\mathbf{\Theta}\in M^+$ is maintained during each iteration as long as the starting value is positive definite, for the same reason that the positive definite constraint is satisfied in the Bayesian fglasso sampler. The full Gibbs sampler is summarized in Algorithm \ref{algorithm:Bayeshorseshoe_Gibbs}. In the case of different truncated number of principal components for each node, the algorithm is straightforward. The block $\mathbf{\Theta}_{ij}$ is rectangle and $(\lambda_{ij}^2)^{(l + 1)}\sim~ \mbox{inverse Gamma}(\frac{M_iM_j+1}{2},\frac{1}{\nu_{ij}^{(l)}}+\frac{\|\mathbf{\Theta}_{ij}^{(l+1)}\|^2_F}{2(\tau^2)^{(L)}})$ and ($\tau^2)^{(l + 1)}\sim \mbox{inverse Gamma}\left(\frac{\sum_{i < j}M_iM_j + 1}{2}, \frac{1}{\zeta^{(l)}}+\sum_{i<j}\frac{\|(\mathbf{\Theta}_{ij})^{(l+1)}\|_F^2}{2(\lambda_{ij}^2)^{(l+1)}}\right).$

\begin{algorithm}[tb]
\SetAlgoLined
\textbf{Input:} Sum of the products matrix $\bm{S}$, i.e., $\bm{S} = \bm{Y}^T\bm{Y}$.\\
\textbf{Output:} Samples of precision matrix $\hat{\bm{\Theta}}$.\\ 
 \textbf{Initialization}: Set $p$ to be number of nodes in graph, set initial values $\bm{\Theta} = \bm{I}$, $\bm{\Sigma} = \bm{I}$, $\bm{\Lambda} = \bm{J}$, $\bm{V} = \bm{J}$, where $\bm{I}$ is $pM\times pM$ identity matrix and $\bm{J}$ is a $pM\times pM$ matrix with all elements equal to one\;
 \While{Given the current $\bm{\Theta}\in M^+$ and $\bm{\tau}$, repeat for a large number of iterations until convergence is achieved}{
    \For {$i = 1,\dots, p$}{
    Partition $\bm{\Theta}$, $\bm{S}$, $\bm{T}$ and $\bm{V}$ into $p\times p$ blocks (focus on updating $i$th column block of $\bm{\Theta}$ corresponding node $i$ and all the other nodes)\;
    1. \For {$j = 1,\dots, M$}{
            
        (1) Partition $\bm{\Theta}^{(l)}$, $\bm{S}$, $\bm{\Lambda}^{(l)}$ and $\bm{V}^{(l)}$ as \eqref{eq:partion_horseshoe}\;
        (2) Draw $\gamma^{(l+1)}\sim~ \mbox{Gamma}\left(\frac{n}{2}+1, \frac{s_{22}+(\lambda_{22}^{(l)})^2}{2}\right)$\;
        (3) Draw $\bm{\beta}^{(l+1)} \sim ~N_{M(p-1)}(-\bm{C}^{(l)}\bar{\bm{s}}_{21}, \bm{C}^{(l)})$, where $\bm{C}^{(l)} = ((\bm{D}_{\tau}^{(l)})^{-1}+(s_{22}+(\lambda_{22}^{(l)})^2)(\overline{\bm{\Theta}_{11}^{-1}})^{(l)})^{-1}$\;
        (4) Update $\bm{\theta}_{21}^{(l+1)} = (\bm{\beta}^{(l+1)},\bm{0}), \bf{\theta}_{12}^{(l)} = \bm{\theta}_{21}^T, \theta_{22}^{(l)} = \gamma^{(l+1)}+(\bm{\beta}^{(l+1)})^T(\overline{\bm{\Theta}_{11}^{-1}})^{(l+1)}\bm{\beta}^{(l+1)}$
    }
    2. Update $\bm{\Lambda}^{(l)}$, i.e., draw sample ($\lambda_{ij}^2)^{(l)}\sim~ \mbox{inverse Gamma}\left(\frac{M^2+1}{2}, \frac{1}{\nu_{ij}^{(l)}}+\frac{\|\bm{\Theta}_{ij}^{(l+1)}\|^2_F}{2(\tau^2)^{(L)}}\right)$\;
    3. Update $\bm{V}^{(l)}$, i.e., draw sample $\nu_{ij}^{(l)}\sim \mbox{inverse Gamma}\left(1,1+\frac{1}{(\lambda_{ij}^2)^{(l+1)}}\right)$\;
    4. Update $\tau^{(l)}$ and $\zeta^{(l)}$, i.e., ($\tau^2)^{(l)}\sim \mbox{inverse Gamma}\left(\frac{M^2(p-1)p+2}{4}, \frac{1}{\zeta^{(l)}}+\sum_{i<j}\frac{\|(\bm{\Theta}_{ij})^{(l+1)}\|_F^2}{2(\lambda_{ij}^2)^{(l+1)}}\right)$, $
\zeta^{(l)}\sim \mbox{inverse Gamma}\left(1,1+\frac{1}{(\tau^2)^{(l+1)}}\right)$\;
    }
    Store the sample precision matrix $\bm{\Theta}$\;
    Increment $l \leftarrow l+1$.
 }
 \caption{Bayesian functional graphical horseshoe Gibbs sampler}\label{algorithm:Bayeshorseshoe_Gibbs}
\end{algorithm}

\subsection{Bayesian FDR-based Inference and Confidence Regions}\label{sec:fdr}
Our goal is to identify significant conditional dependence between different nodes, which can subsequently be mapped into edges in the estimated graph. An intuitive way is to identify blocks with the Frobenius norm of block at least $\delta$, which could be a practical threshold. We consider the direct posterior probability approach; i.e., Bayesian FDR-based inference \citep{storey2003positive, morris2011automated} to threshold in a way that considers both statistical and practical significance.

First, we compute the edge strength as $\|\hat{\mathbf{\Theta}}_{ij}\|_F$ by \eqref{eq:fedge} for $i, j = 1, \dots, p$ for each sample of the precision matrix. Then we define a threshold $\delta$ as practical significance, which could be determined by or associated with some prior knowledge such as the desired sparsity of graph. For example, the value of $\delta$ for a desired sparsity level of 95\% will be higher than the one for the desired sparsity level is 90\%. Further, we can estimate the posterior probability of $\|\hat{\mathbf{\Theta}}_{ij}\|_F$ at least $\delta$ intensity as 
\begin{equation}
\label{eq: Bayesian fdr}
 p_{ij}^{\delta} = \mbox{Pr}\{\|\hat{\mathbf{\Theta}}_{ij}\|_F > \delta\} \approx \sum_{l = 1}^L\frac{1}{L}I\{\|\hat{\mathbf{\Theta}}^{l}_{ij}\|_F > \delta\}   
\end{equation}
for $i, j = 1, \dots, p$, and $L$ is length of MCMC output. The quantity $1 - p_{ij}^{\delta}$ can be considered as a natural ``Bayesian posterior p-value" or ``positive false discovery rate" analogue of p-value (also named ``q-value" by \citep{storey2003positive}). Given a significance level $\alpha$, we then identify the significant blocks by $E = \{
(i, j): p_{ij}^{\delta} > \phi_{\alpha}^{\delta} \}$, where $\phi_{\alpha}^{\delta}$ is a threshold on the posterior probabilities that controls the average Bayesian FDR at level $\alpha$. Following Morris et al.\citep{morris2011automated}, we sort the $p_{ij}^{\delta}$ in descending order to yield $p_{(ij)}^{\delta}$. Then
\begin{equation}
   \phi_{\alpha}^{\delta} = p^{\delta}_{(i^*j^*)}, 
\end{equation}
where $(i^*, j^*) = \argmax\{(i, j): \frac{1}{B}\sum_{{(ij)}}(1 - p_{(ij)}^{\delta})$. 

Another method to identify significant blocks among blockwise precision matrices is via credible regions. Let $\mathbf{\hat{\theta}}_{ij}$ be the vectorized representation of $\mathbf{\hat{\Theta}}_{ij}$. For each edge, we have $\mathbf{\hat{\theta}}_{ij}^{(1)}, \dots, \mathbf{\hat{\theta}}_{ij}^{(L)}$ posterior samples, then we can calculate the sample covariance $\Sigma_{\mathbf{\theta}_{ij}}$. The approximate $(1 - \alpha) \times 100 \%$ joint confidence region for $\mathbf{\hat{\theta}}_{ij}$ is 
\begin{equation}
    R_{ij} = \{\mathbf{\hat{\theta}}_{ij} | (\mathbf{\hat{\theta}}_{ij} - \bar{\theta}_{ij})^T \Sigma_{\mathbf{\hat{\theta}}_{ij}}^{-1}(\mathbf{\hat{\theta}}_{ij} - \bar{\theta}_{ij}) \leq q^*\}
\end{equation}
where $q^*$ is the smallest possible $q$ such that 
\begin{equation}
    \sum_{l = 1}^L I\{(\mathbf{\hat{\theta}}_{ij} - \bar{\hat{\theta}}_{ij})^T \Sigma_{\mathbf{\hat{\theta}}_{ij}}^{-1}(\mathbf{\hat{\theta}}_{ij} - \bar{\theta}_{ij}) \leq q\} \geq L(1 - \alpha).
\end{equation}
The approximate volume of confidence regions with level $(1-\alpha) \times 100\%$ is proportional to $\chi_{p'\alpha}^2|\Sigma_{\hat{\theta}_{ij}}|^{\frac{1}{2}}$, where $p'$ is the dimension of $\hat{\theta}_{ij}$.

\section{Numerical Experiments}\label{sec:sims}
We designed simulation studies to assess the performance of both our proposed Bayesian functional graphical lasso and the proposed horseshoe outlined in Section \ref{sec:prop}. For our simulation studies, we considered different sample sizes ($N = 5, 20, 100, 200$), graph sizes ($p = 10, 30, 50$), two different types of networks, and both sparse- and dense-sampled functions. This allows us to compare the frequentist fglasso\citep{qiao2019functional}, Bayesian fglasso, and functional graphical horseshoe to each other across a variety of scenarios. We assess classification accuracy and fidelity of the estimates of both the zero and non-zero entries of the precision matrices in Subsection \ref{subsec:performance}. 



\subsection{Estimation and Classification}\label{subsec:performance}
Similar to the simulation studies considered by \cite{qiao2019functional}, we simulate functional data with $g_{ij}(t) = \mathbf{s}(t)^T\B{\delta}_{ij}, ~i=1, \ldots, n, ~j= 1, \ldots, p$, where $\mathbf{s}(t) \in L(\mathcal{T})^5$ contains the first five Fourier basis functions, and $\B{\delta}_{ij}\in \mathbb{R}^5$ is a zero mean Gaussian random vector. Hence, $\B{\delta}_i = (\B{\delta}_{i1}^T,\dots,\B{\delta}_{ip}^T)^T \in \mathbb{R}^{5p}$ follows a multivariate Gaussian distribution with covariance matrix $\mathbf{\Sigma}=\mathbf{\Theta}^{-1}$, where the underlying graph is determined by the sparsity pattern of $\B{\Theta}$. We consider here two types of networks:
\begin{itemize}
    \item {\em Network 1}: A block banded matrix $\mathbf{\Theta}$ with $\mathbf{\Theta}_{jj} = \mathbf{I}_5$, $\mathbf{\Theta}_{j,j-1} = \mathbf{\Theta}_{j-1,j} = 0.4\mathbf{I}_5$, and $\mathbf{\Theta}_{j,j-2} = \mathbf{\Theta}_{j-2,j} = 0.2\mathbf{I}_5$ for $j = 1,\cdots, p$, and 0 elsewhere. The network results in each node being connected to its immediate neighbors, and weaker connection to its second-order neighbors. \\
    
    \item {\em Network 2}: For $j=1,\dots, 10, 21, \dots, 30, \dots$, the corresponding submatrices in $\mathbf{\Theta}$ are the same as those in Network 1 with $p = 10$, indicating every alternating block of 10 nodes are connected as Network 1. For $j = 11, \dots, 20, 31, \dots, 40, \dots,$ we set $\mathbf{\Theta}_{jj} = \mathbf{I}_5$, so the remaining nodes are fully isolated.
\end{itemize}

For each network, we generate $n$ realizations of $\B{\delta} \sim N_{5p}(\B{0}, \B{\Theta}^{-1})$. The observed data are then generated as $h_{ijk} = g_{ij}(t_{ik})+e_{ijk},\quad e_{ijk}\sim N(0, 0.5^2), ~~k= 1, \ldots, T$, where subject $i$ was observed at time points $t_{i1},\dots, t_{iT} \in [0, 1]$. We consider two sampling schemes from the functions:
\begin{itemize}
    \item {\em Dense design with equally spaced measurements}: Each function was recorded on a regular grid between 0 and 1, i.e., $t_{i1}=0, \dots, t_{iT}=1$ and $T= 100, ~i = 1,\dots, N$.\\
    
    \item {\em Sparse design with irregularly-spaced measurements}: Each function was recorded randomly; i.e., $t_{ik}$ are drawn randomly between 0 and 1 for $k = 1, \dots, 9$. 
\end{itemize}

Our proposed graphical models work with any choice of basis representation, but we choose the data-driven functional PCA approach, due to the mean-square optimality discussed in Section \ref{sec: FPCA} and the smoothness of the simulated data. Thus, to implement any of the approaches considered, we need to compute the first $M$ estimated principal components scores of $g_{ij}$. We use the PACE algorithm \citep{yao2005functional} for the irregularly sampled setting via the \texttt{fdapace} package in \texttt{R} \citep{fdapacePkg}. For the regularly sampled case, we use ordinary singular value decomposition. We determine the truncation level $M$ in \eqref{eq:m-truncated} using the minimum number of principal component to capture 95\% of the variability over all nodes for the SVD method and PACE algorithm. For both the Bayesian functional graphical lasso and the Bayesian functional graphical horseshoe, a total of 10,000 MCMC iterations are performed after 1000 burn-in iterations. Convergence is assessed via trace plots of randomly selected elements of $\B{\Theta}$.  

We first compare the frequentist fglasso to our proposed Bayesian fglasso model with a fixed regularization parameter $\lambda$. The primary differences then are the point estimates of $\B{\Theta}$ (MAP estimate versus posterior mean) and the thresholding rule used to select the graph. With MAP estimation, the zeros are automatically produced as part of the optimization. For the Bayesian procedure, elementwise equal-tailed credible intervals are constructed from the MCMC output, whence elements of $\B{\Theta}$ are selected via those intervals that do not contain zero. For both methods and for a grid of $\lambda$ values, we compute the true positive rate, $TPR_{\lambda} = TP/(TP+FN)$, and false positive rate, $FPR_{\lambda} = FP/(FP+TN)$, where $TP$ and $TN$ stand for true positives and negatives, respectively in terms of network edges correctly identified, and similarly for $FP$ and $FN$. With these measures we can compute the areas under the associated ROC curves (AUCs), where values closer to 1 indicate better discriminative power between the true zero and non-zero edges in the true graph.

Table \ref{tab: auc_fglasso} displays the AUC values for the various settings described above. As expected from the correspondence between the frequentist and Bayesian formulations of the objective, we see very similar performance between the two approaches across each scenario considered. This suggests that, as far as graphical model selection is concerned, the frequentist and Bayesian implementations of the fglasso have equivalent discriminative ability with fixed a regularization parameter. However, the Bayesian approach facilitates additional flexibility in how the regularization parameter is treated. While it is possible to estimate $\lambda$ with, e.g., cross-validation, it can be computationally expensive to do so. On the other hand, assigning a prior distribution to the regularization term, or circumventing an explicit $\lambda$ altogether via, e.g., the horseshoe, allows the appropriate penalty to be learned from the data along with the remaining parameters in the model, so that it only needs to be fit once. 
\begin{table}[!tb]
    \begin{minipage}{0.5\linewidth}
      \centering
    \begin{tabular}{l l l l}
        \multicolumn{4}{c}{Network 1, $p = 10$, dense data}\\
        & N = 20      & N = 100     & N = 200     \\ \hline
        Bayes & 0.65 (0.10) & {\bf 0.88} (0.04) & {\bf 0.94} (0.03) \\
         FGM   & {\bf0.66} (0.06) & 0.84 (0.03) & 0.90 (0.03) \\\hline
         & & &\\
         \multicolumn{4}{c}{Network 1, $N = 100$, dense data}\\
             & p = 10      & p = 30      & p = 50      \\ \hline
             Bayes & {\bf 0.88} (0.04) &  {\bf 0.86} (0.02) &   0.83 (0.02)  \\
            FGM   & 0.84 (0.03) & 0.84 (0.02) &  {\bf 0.85} (0.02)    \\ \hline
    \end{tabular}
    \end{minipage}%
    \begin{minipage}{0.5\linewidth}
      \centering
    \begin{tabular}{l l l}
        \multicolumn{3}{c}{Network 1, $p = 10$, N = 100}\\
        & Dense       & Sparse      \\ \hline
        Bayes & {\bf 0.88} (0.04) & {\bf 0.83} (0.04) \\
        FGM   & 0.84 (0.03) & 0.80 (0.04) \\ \hline
         & & \\
         \multicolumn{3}{c}{$p= 10$, $N = 100$, dense data}\\
             & Network 1   & Network 2   \\ \hline
             Bayes & {\bf 0.88} (0.04) & 0.91 (0.04) \\
            FGM   & 0.84 (0.03) & {\bf 0.92} (0.04) \\ \hline
    \end{tabular}
    \end{minipage} 
\caption{The mean area under the ROC curves for Bayesian fglasso and FGM. Standard errors are shown in parentheses.}
\label{tab: auc_fglasso}
\end{table}

Next we compare the frequentist fglasso to the hierarchical Bayesian fglasso (as opposed to the fixed-$\lambda$ Bayesian fglasso considered above) and the functional graphical horseshoe in terms of misclassification error and graph similarity when the regularization parameter is determined as it would be in practice. The frequentist fglasso requires tuning of the regularization parameter, either through cross-validation or by setting it to yield a desired sparsity level. For the frequentist fglasso we use 10-fold cross-validation for model selection, since in practice the { true} sparsity of the underlying graph will most likely be unknown. In frequentist fglasso, the sparsity of the estimated graph could be tuned by the parameter $\lambda$. In the Bayesian framework, the size of the credible intervals could also adjusted to desired sparsity. As the Bayesian fglasso behaves differently than the horseshoe, and to reduce the number of false positives, we use 90\% credible intervals for the hierarchical Bayesian fglasso. For the functional graphical horseshoe, we follow \cite{li2019graphical} and use 50\% credible intervals for thresholding. Table \ref{tab: comp_all} reports the false positive rate (FPR), false negative rate (FNR), overall misclassification rate, the F1 score (also known as Dice coefficient), and resulting sparsity levels for $p = 10, 30, 50$ nodes, and $N = 100$ subjects with densely-sampled functions under both Network 1 and 2. The scores are averaged over 10 replications of each scenario. With the exception of $p=10$ nodes in Network 2, the functional graphical horseshoe always has the highest F1 score, indicating the strongest graph similarity. We further see that the functional graphical horseshoe has the results with closest sparsity level to the true sparsity level. The frequentist fglasso generally performs the worst, due to the overly dense graphs that it produces, reflecting known risks of using log-likelihood-based cross-validation in this type of model \citep{wasserman2009high}. 

\begin{table}[tb]
\footnotesize
\centering
\begin{tabular}{cccccccc}
\hline
                                  &               & \multicolumn{3}{c}{\textit{\textbf{Network 1}}} & \multicolumn{3}{c}{\textit{\textbf{Network 2}}} \\ \hline
                                  &               & p=10            & p=30          & p=50          & p=10            & p=30          & p=50          \\ \hline
\multirow{5}{*}{\rotatebox[origin=c]{90}{\textbf{fglasso}}}  & FPR (\%)      & 6.07 (3.59)     & 0.95 (0.46)   & 13.87 (0.65)  & 8.61 (4.38)     & 1.13 (0.53)   & 16.06 (1.05)  \\
                                  & FNR (\%)      & 38.82 (6.00)    & 62.46 (5.61)  & 32.27 (4.45)  & 24.44 (9.69)    & 50.00 (3.80)  & 27.78 (5.69)  \\
                                  & ERR (\%)      & 18.44 (3.73)    & 9.01 (1.01)   & 15.33 (0.67)  & 11.78 (4.34)    & 4.16 (0.48)   & 16.41 (1.06)  \\
                                  & F 1           & 0.71  (0.06     & 0.52  (0.06)  & 0.41  (0.02)  & $\bm{0.72  (0.09)}$    & 0.60  (0.04)  & 0.21  (0.02)  \\
                                  & Sparsity (\%) & 26.89           & 5.75          & 18.14         & 22              & 4.16          & 17.71         \\ \hline
\multirow{5}{*}{\rotatebox[origin=c]{90}{\textbf{fghorse}}}  & FPR (\%)      & 18.57 (8.27)    & 2.14 (0.60)   & 1.03 (0.36)   & 10.83 (6.97)    & 0.74 (0.64)   & 0.40 (0.21)   \\
                                  & FNR (\%)      & 22.35 (9.41)    & 36.14 (3.35)  & 41.81 (4.18)  & 24.44 (9.69)    & 41.11 (3.08)  & 43.06 (3.11)  \\
                                  & ERR (\%)      & 20.00 (6.13)    & 6.60 (0.78)   & 4.26 (0.37)   & 13.56 (6.16)    & 3.24 (0.52)   & 1.66 (0.23)   \\
                                  & F 1           & $\bm{0.75 (0.08)}$     & $\bm{0.72 (0.03)}$   & $\bm{0.68 (0.03)}$   & 0.70 (0.11)     & $\bm{0.69 (0.03)}$   & $\bm{0.67 (0.04)}$   \\
                                  & Sparsity (\%) & 40.89 (6.38)    & 10.23 (0.57)  & 5.56 (0.54)   & 23.78 (5.63)    & 4.34 (0.72)   & 2.07 (0.22)   \\ \hline
\multirow{5}{*}{\rotatebox[origin=c]{90}{\textbf{FGM}}}      & FPR (\%)      & 81.07  (10.11)  & 44.58 (4.48)  & 31.35 (4.08)  & 50.00 (16.76)   & 19.39 (6.44)  & 11.14 (2.22)  \\
                                  & FNR (\%)      & 4.71 (6.86)     & 12.28 (2.48)  & 19.28 (4.68)  & 1.11 (3.33)     & 17.04 (5.79)  & 19.72 (5.19)  \\
                                  & ERR (\%)      & 52.22 (5.28)    & 40.34 (3.85)  & 30.39 (3.44)  & 40.22 (13.42)   & 19.24 (5.87)  & 11.40 (2.10)  \\
                                  & F 1           & 0.58  (0.03)    & 0.36  (0.02)  & 0.30  (0.02)  & 0.51   (0.09)   & 0.36  (0.06)  & 0.30  (0.04)  \\
                                  & Sparsity (\%) & 86.44           & 50.23         & 35.26         & 59.78           & 23.33         & 13.18         \\ \hline
\multicolumn{2}{l}{True sparsity (\%)}            & 37.78           & 13.1          & 7.92          & 20              & 6.21          & 2.94          \\ \hline
\end{tabular}
\caption{Summary statistics of false positive (FPR), false positive rate (FNR), mislassification rate (ERR), F1 (Dice) score and estimated graph sparsity of graph estimation with 10 data sets generated by dense functional data with underground Network 1 and 2 separately. ``fglasso" refers to the Bayesian functional graphical lasso method with gamma prior for shrinkage parameter $\lambda$; ``fghorse" refers to the functional graphical horseshoe method, while "FGM" refers to the frequentist version of functional graphical lasso model proposed by \citet{qiao2019functional}. The means are reported here, the standard errors are shown in paratheses.}
\label{tab: comp_all}
\end{table}

We turn our attention to a direct comparison between the Bayesian fglasso and the functional graphical horseshoe. One advantage of the horseshoe prior is that the global/local shrinkage in the model automatically adapts to the observed data. To allow for such ``automatic" adaptation in the Bayesian fglasso, we use the augmented version of the model in which a hyperprior is assigned, $\lambda^2\sim \mbox{Gamma}(1, 0.01)$. In this comparison, we consider fpc scores generated from $p = 10$ nodes with $N = 100$ observations and rank $M = 5$ in the basis expansion, using Network 1 defined above. For evaluation, we construct credible regions for each model. Given confidence level, the volume of credible regions is proportional to $|\Sigma_{\hat{\theta}_{ij}}|^{\frac{1}{2}}$. Performance characteristics are quantified in Table \ref{tab: lasso_vs_horse}, where normalized logarithms of $|\Sigma_{\hat{\theta}_{ij}}|^{\frac{1}{2}}$ are reported. In addition, Bayesian FDR $ p^{\delta}_{ij}$ are calculated, where $1 - p^{\delta}_{ij}$ can be considered ``q-values", or estimates of the local false discovery rate \citep{storey2003positive}. 

\begin{figure}
    \centering
    \includegraphics[width = 0.6\textwidth]{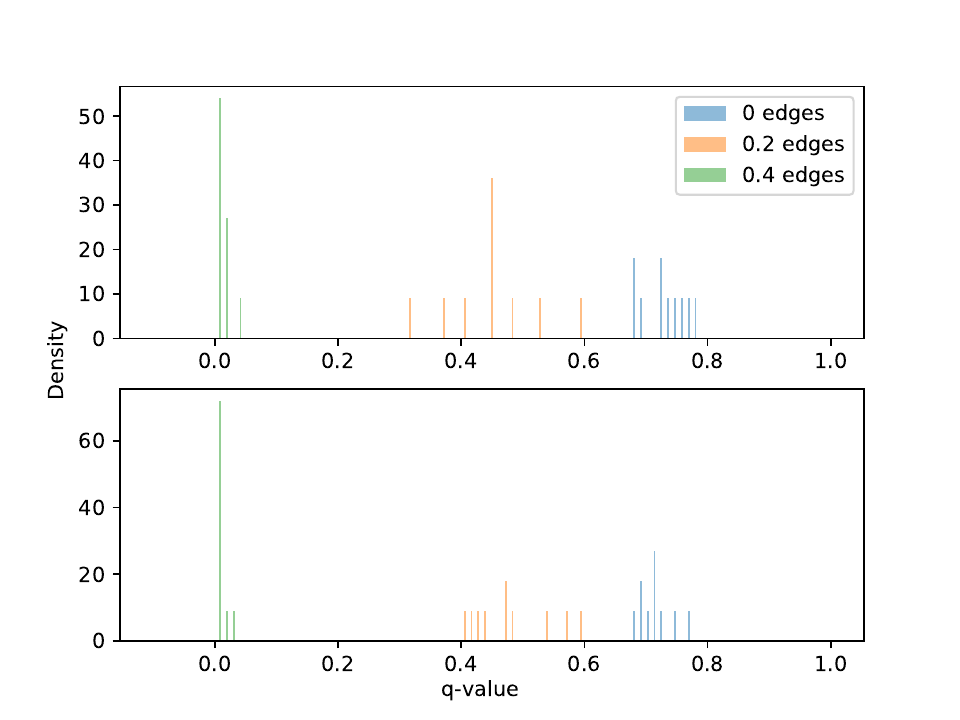}
    \caption{Histograms of q-values for Bayesian functional graphical lasso (top) and functional graphical horseshoe (bottom)}
    \label{fig: Bayes_com}
\end{figure}

Figure \ref{fig: Bayes_com} depicts the Bayesian FDR-based ``q-value" for the off-diagonal blocks, separated by the zeros, those with edge weight 0.2, and those with edge weight 0.4. We use the identical practical significance threshold $\delta$ (60\% quantile of off-diagonal $\|\hat{\mathbf{\Theta}}_{ij}\|_F$) for both Bayesian functional lasso and horseshoe. We can make several observations from this figure. First, the Bayesian fglasso results in more false positives, compared to the lower false positive rate with the same rule applied to the functional graphical horseshoe. Thus, the Bayesian functional graphical lasso exhibits behavior similar to that which is known about its scalar counterpart. The stronger mass near the origin applied by the horseshoe compared to the Bayesian graphical lasso\citep{li2019graphical} results in much better identification of the true zeros in the model. Second, for the truly non-zero entries in $\B{\Theta}$, the functional graphical horseshoe q-values show stronger separation compared with the Bayesian fglasso. In Table \ref{tab: lasso_vs_horse} we see the volume of the functional graphical horseshoe credible regions growing as the signal size goes from 0.2 to 0.4, as would be expected from the results of \cite{VanDerPasEtAl14}. For the non-zero estimates, we observe more uncertainty under the Bayesian functional graphical horseshoe as the signal size grows from 0.2 to 0.4, while fglasso has less uncertainty as the signal strength increases. Hence, edge selection via Bayesian FDR / credible regions under the horseshoe tends to be conservative and that the volumes of the credible regions tends to grow with size of the true signal \citep{VanDerPasEtAl14}.

\begin{table}
\centering
\begin{tabular}{lrrr} 
        &0 edges& 0.2 edges & 0.4 edges  \\ 
\hline
fglasso & 0.304 (0.056)                  & 0.019 (0.079)                     & -0.962 (0.150)                     \\
fghorse & -0.579 (0.037)                  & 0.169 (0.152)                     & 1.650 (0.060)                     \\
\end{tabular}
\caption{Summary of credible region volumes (log scale) for Bayesian functional glasso and horseshoe based on 10 replications}
\label{tab: lasso_vs_horse}
\end{table}

\section{Applications}\label{sec:apps}
\subsection{Alcoholism Study}\label{sec:EEG}
Here we apply our proposed functional graphical horseshoe method to an event-related potential electroencephalography (EEG) dataset from an alcoholism study\citep{zhang1995event}. The data, freely available at \url{https://archive.ics.uci.edu/ml/datasets/eeg+database}, consist of 122 subjects, 77 of whom were identified as alcoholics, and 45 in the control group. Signals were initially collected from 64 electrodes placed on subjects' scalps at standard positions, and captured voltage signals at 256 Hz during a one-second time period. 120 trials were collected per subject. During each trial, the subject was exposed to either a single stimulus (a single picture) or two stimuli (a pair of pictures) shown on a computer monitor. The interest here is in estimating a network representing functional connections between different brain regions. We filtered the signals through a banded filter between to obtain $\alpha$ frequencies between 8 and 12.5 Hz, as the $\alpha$ band has been shown to differentiate between alcoholic and control groups under this task. Moreover, we take the average of all trials for each subject, resulting in a single event-related potential curve at each electrode for each subject. The filtering was performed by applying the \texttt{eegfilter} function in the \texttt{eegkit} package \citep{helwig2018eegkit} in \texttt{R}. Since the data are regularly and densely sampled, we employed the regular SVD method to compute the principal component scores $\hat{\bf{a}}$. The truncated number of fpcs are selected so that at least 95\% of the variation in the filtered signal trajectories for control and alcoholic curves are captured by the basis representations. For each MCMC run, 10,000 iterations were retained after an initial burn-in period of 1000 iterations. To be conservative and achieve a higher level of sparsity, we are interested in finding the edges through Bayesian FDR-based thresholding.

\begin{figure}
\centering
\begin{subfigure}{\textwidth}
  \includegraphics[width=1\linewidth]{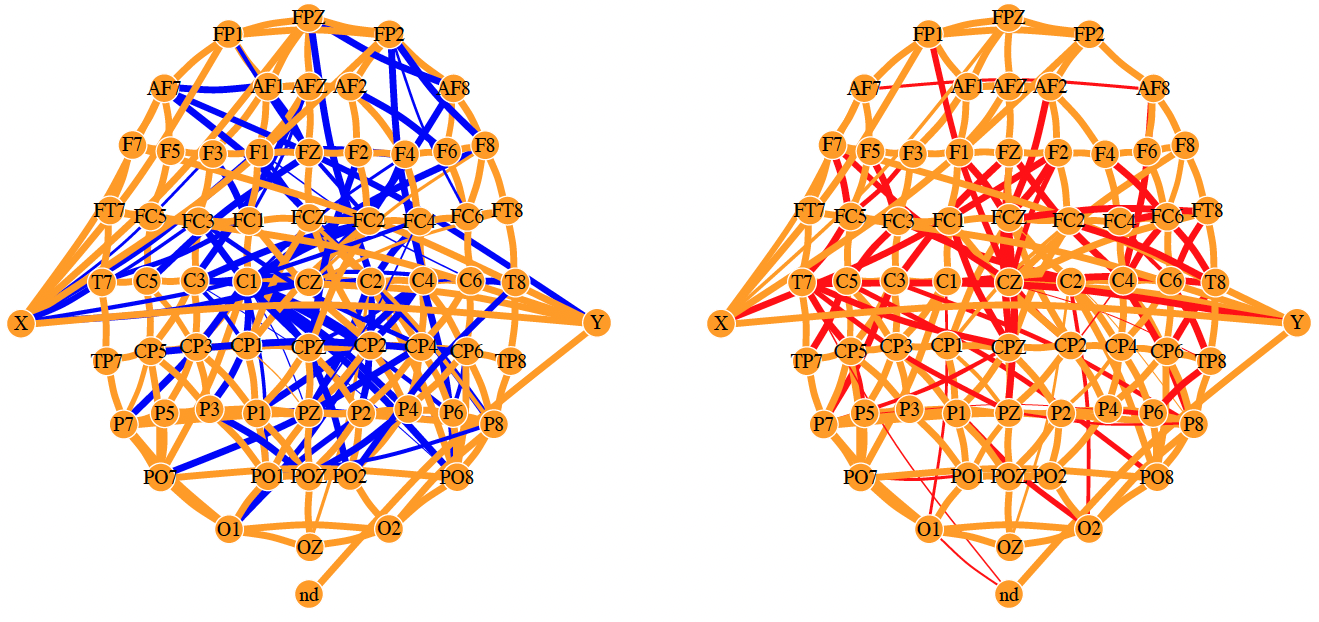}
\end{subfigure}
\caption{Functional brain connectivities for the alcoholic (left) and control (right) groups constructed by Bayesian functional graphical Horseshoe by controlling false discovery rate $\alpha = 0.01$.}
\label{fig:eeg_weight}
\end{figure}

The results are summarized in Figure \ref{fig:eeg_weight}. The weight of each edge is evaluated as $p_{ij}^{\delta} = \mbox{Pr}\{\|\hat{\mathbf{\Theta}}_{ij}\|_F > \delta\} $. The practical threshold $\delta$ is selected to be the 60\% percentile of $\|\hat{\mathbf{\Theta}}_{ij}\|_F$ and $\alpha$ is 0.01 in Figure \ref{fig:eeg_weight}. In the weighted graphs, thicker edges indicate larger weights. We can see that most of the common edges have strong weights. By contrasting the alcoholic and control graphs, we see that the alcoholic group contains more edges connecting the frontal-central regions than the control group. It also appears that the right parietal region tends to have more connection in the alcoholic group than in the control group. Finding a more densely connected frontal region and differences in the right parietal region agrees with that which was found by \cite{zhu2016bayesian}. There are clear differences, of course, just as there were between functional graphical models estimated by \cite{qiao2019functional} and \cite{zhu2016bayesian}. Overall, however, we find qualitative agreement with the analysis of \cite{zhu2016bayesian}, despite the fact that their assumed model was quite different from our functional graphical horseshoe, and the fact that our approach does not assume a decomposable graph. 

\subsection{Structural Connectivity after Traumatic Brain Injury}\label{sec:ADNI}
It is known that traumatic brain injuries can cause acute disconnections in white matter tracts \citep{RutgersEtAl08}, and there is interest in studying what happens to these connections during the chronic phase after such injuries. To address this question, we applied our proposed Bayesian functional graphical horseshoe to diffusion tensor imaging data (DTI) obtained from the Alzheimer's Disease Neuroimaging Initiative (ADNI). The data consist of 34 subjects, 17 of whom have been identified as having experienced traumatic brain injury (TBI) with the remaining 17 being healthy controls. The control subjects were selected from a much larger group via propensity score matching to control for confounding variables such as patient's age, sex, whole brain volume, and Alzheimer's disease status. The data contain initial and follow-up measurements of fractional anisotropy in each of $p=26$ regions of interest (ROI) in the brain, resulting in irregularly measured longitudinal data for each ROI. For each subject, anywhere from one to nine time points are available, each separated by several months. The data preprocessing includes eddy-correction \citep{andersson2016integrated}, brain extraction, \citep{smith2002fast} and intensity normalization \citep{jenkinson2001global,jenkinson2002improved}. For each voxel in a brain image, the fractional anisotropy (FA) is calculated as $FA = (3/2)((\lambda_1-\Bar{\lambda})^2+ (\lambda_2-\Bar{\lambda})^2+(\lambda_3-\Bar{\lambda})^2)^{1/2}(\lambda_1^2+\lambda_2^2+\lambda_3^2)^{-1/2}$,
where $\lambda_1, \lambda_2$ and $\lambda_3$ are the eigenvalues associated with the $x-$, $y-$ and $z-$ directions of the diffusion tensor and $\Bar{\lambda} = \frac{\lambda_1+\lambda_2+\lambda_3}{3}$ is the mean diffusivity. FA indicates the degree of anisotropy of a diffusion process valued between 0 and 1. If the FA is close to 0, the diffusion is unrestricted in all directions, indicating loosely structured (i.e., deteriorated) white matter, whereas FA close to 1 means that the diffusion mainly occurs only along one axis, thus indicating stronger white matter in that area. The voxel-wise FA values are averaged to summarize the observed FA in each ROI at each time point. In total, we focus on exploring the connectivity (conditional dependence in brain atrophy) between 52 ROIs. Each ROI considered is listed in Table \ref{table:ADNI_ROI}.
\begin{table}[tb]
\centering
\begin{tabular}{p{0.07\linewidth}|p{0.15\linewidth}|p{0.6\linewidth}} 
\hline
\textbf{Label}  &\textbf{Hemisphere}& \textbf{ROI}                                                        \\ 
\hline
CST             &left, right &Corticospinal tract                                  \\ 
ICP             &left, right & Inferior cerebellar peduncle                         \\ 
ML              &left, right &Medial lemniscus                                      \\ 
SCP             &left, right &Superior cerebellar peduncle                        \\ 
CP              &left, right &Cerebral peduncle                                     \\ 
ALIC            &left, right &Anterior limb of internal capsule                    \\ 
PLIC            &left, right &Posterior limb of internal capsule                   \\ 
PTR             &left, right &Posterior thalamic radiation, includes optic radiatio\\ 
ACR             &left, right &Anterior corona radiata                               \\ 
SCR             &left, right &Superior corona radiata                               \\ 
PCR             &left, right &Posterior corona radiata                              \\ 
CGC             &left, right &Cingulum, cingulate gyrus                              \\ 
CGH             &left, right &Cingulum (hippocampus), cingulate gyrus                                \\ 
FX\_ST          &left, right &Fornix (cres) / Stria terminalis                     \\ 
SLF             &left, right &Superior longitudinal fasciculus                    \\ 
SFO             &left, right &Superior fronto-occipital fasciculus, could be a part of anterior internal capsule     \\ 
IFO             &left, right &Inferior fronto-occipital fasciculus                \\ 
SS              &left, right &Sagittal stratum, includes inferior longitudinal fasciculus and
inferior fronto-occipital fasciculus                                                 \\ 
EC              &left, right &External capsule                                 \\ 
UNC             &left, right &Uncinate fasciculus                                  \\ 
FX              &left, right &Fornix, column and body of fornix                                    \\ 
GCC             &left, right &Genu of corpus callosum                              \\ 
BCC             &left, right &Body of corpus callosum                              \\ 
SCC             &left, right &Splenium of corpus callosum                        \\ 
RLIC            &left, right &Retrolenticular part of internal capsule             \\ 
TAP             &left, right &Tapatum                                             \\ 
\hline
\end{tabular}
\caption{Index of the ``Eve” white matter atlas labels corresponding to Figure \ref{fig:FA_MD_w} the TBI connectivity study.}
\label{table:ADNI_ROI}
\end{table}

Similar to the EEG study, we assume the longitudinal data $g_1(t), \dots, g_{52}(t)$ jointly follow a $52$-dimensional MGP. We use the PACE algorithm\citep{yao2005functional} to carry out FPCA using \texttt{fdapace} package \citep{carroll2020fdapace}, where the truncation level of fpc scores is selected by capturing 95\% of the variance in the curves. Then we fit our functional graphical horseshoe model via the Gibbs sampler discussed in \ref{sec:horse}, running the MCMC  for 100,000 iterations after an initial 10,000-iteration burn-in period. The estimated graphs are constructed based on Bayesian FDR-based inference from the approximate posterior sample, at level $\alpha = 0.1$ with practical significance level $\delta$ set to be the $0.60$ quantile of off-diagonal $\|\hat{\mathbf{\Theta}}_{ij}\|_F$. 

\begin{figure}[!t]
\centering
\begin{subfigure}[]{\textwidth}
  \includegraphics[width=1\linewidth]{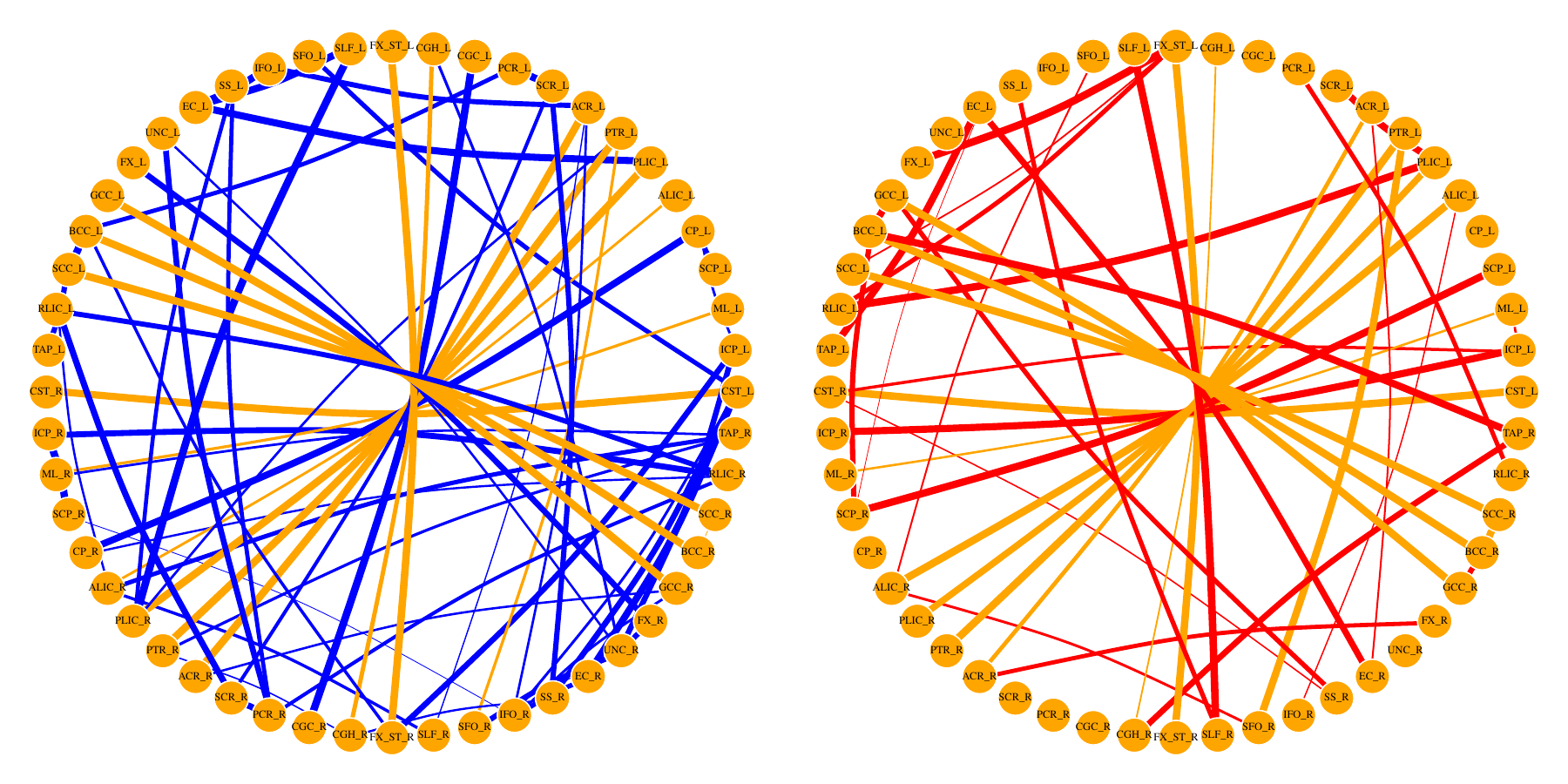}
\end{subfigure}
\begin{subfigure}[]{\textwidth}
  \includegraphics[width=1\linewidth]{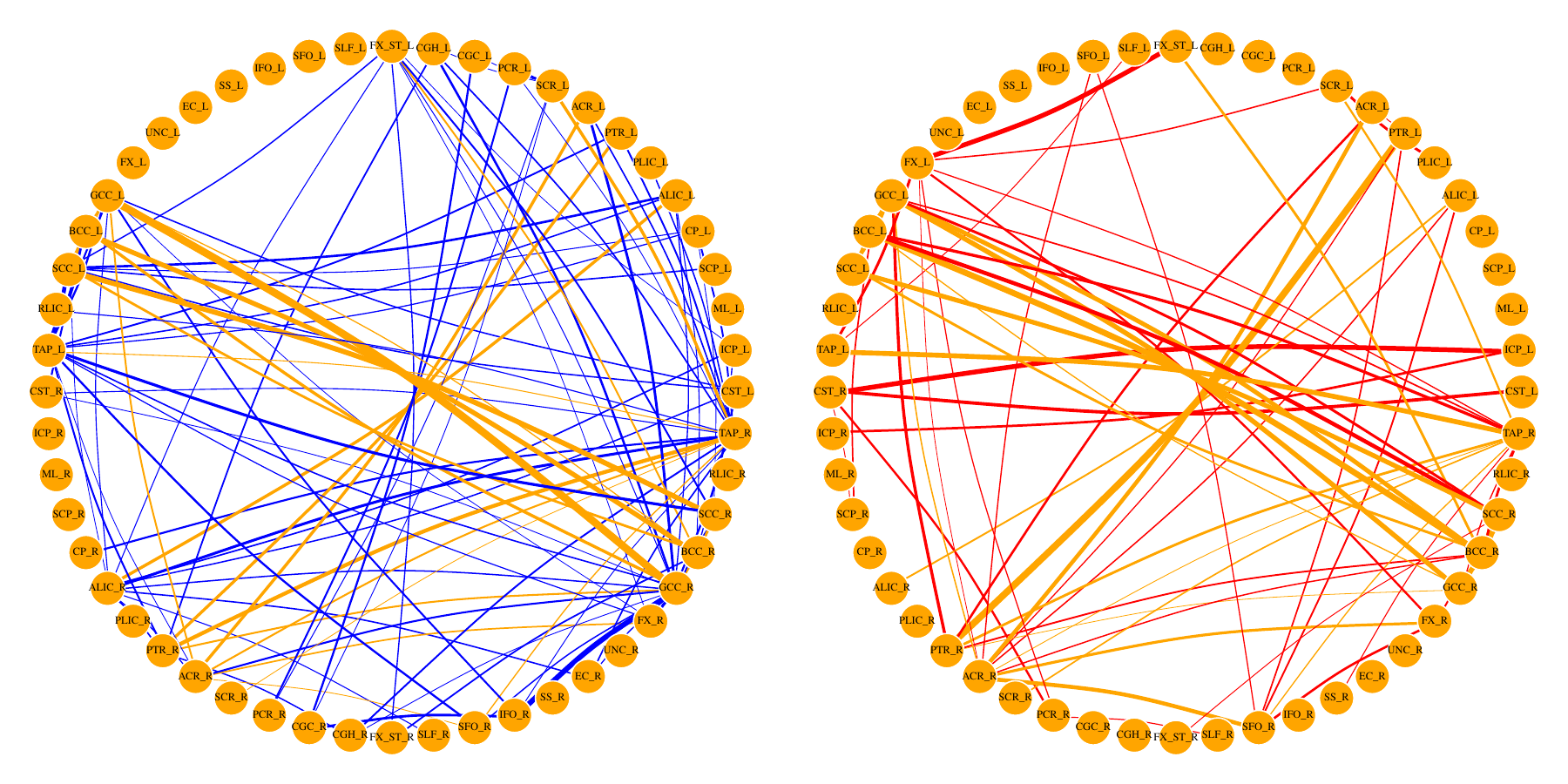}
\end{subfigure}
\caption{The estimated weighted brain connectivity graphs for TBI (left) and control (right) groups by Bayesian horseshoe (upper) and the frequentist fglasso (lower). Nodes names are marked with abbreviation of ROIs, which are defined in Table \ref{table:ADNI_ROI}. The thicker edges indicate larger weights; Blue lines denote edges identified only by the TBI group, red denote edges identified only by the control group, and orange denote edges identified by both groups.} 
\label{fig:FA_MD_w}
\end{figure}

The estimated FA networks for the TBI and control groups are plotted in top panel of Figure \ref{fig:FA_MD_w}. The sparsity levels for the TBI and control groups are 4.60\% (62 edges) and 2.94\% (39 edges), respectively. Twelve edges are common to both graphs, indicated by orange lines in the figure. The FA networks estimated via frequentist FGM\citep{qiao2019functional} for TBI and control groups are also displayed in the bottom panel of Figure \ref{fig:FA_MD_w}. In both results, the TBI group tends to have more connections between different ROIs, particularly within right hemisphere, though the Bayesian horseshoe produces stronger differences in edgeweights than the FGM. The increased connectivity in the right hemisphere observed here may reflect degraded right hemisphere functioning as performance IQ (PIQ) deteriorates. This reflects the fact that in most types of brain damage and dysfunction, PIQ tends to deteriorate faster than verbal IQ, leading to a PIQ/VIQ discrepancy. 

Overall, we see a denser set of connections in the TBI group than in the healthy controls, similar to a phenomenon that has been previously observed in children\citep{KookEtAl20}. The increased structural connectivity observed here in the TBI patients may reflect so-called ``axonal sprouting" that occurs as disconnected neurons attempt to reconnect to a network from which they have been isolated. This results in ``improved" connectivity, but nevertheless a deterioration of functional activation and task performance. It is also possible that there are no entirely new structural connections between areas, but that the weight of the white matter tracts in the areas may increase, on average, because targeted selection of cortical modules necessary to participate in a given task breaks down in the presence of a brain injury. Indeed, enlargement of the activation field in stroke recovery has been observed in the literature \citep{LindowEtAl16}. Our results here support the previously observed phenomenon in which white matter tracts evince increased connectivity in the chronic phase after disconnection of the tracts in the acute phase \citep{RutgersEtAl08}.

Application of our proposed Bayesian functional graphical model estimation procedure to the two studies in this section illustrate both the feasibility of implementation as well as meaningful results that may be produced. The EEG study was able to provide insights similar to previous analyses, but with fewer assumptions on the structure of the underlying graph. The TBI results likewise show qualitative similarity with results obtained under the frequentist approach, with added benefit of having measures of uncertainty about the estimated graph. Further, the stronger differences in estimated edge weights produced under our proposed approach may produce additional insights concerning the degradation of cortical modules after a brain injury.

\section{Conclusion}\label{sec:summary}
In this article, we consider a Bayesian framework for graphical models associated with functional data. We proposed a fully Bayesian version of the  functional graphical lasso as well as a novel functional graphical horseshoe prior. We provide also easily implemented Gibbs samplers via auxiliary variables to induce conditional conjugacy and adapting matrix partitioning techniques that have been used for other MCMC implementations of Bayesian graphical models. We compared these models to each other in several simulated scenarios. The Bayesian fglasso with fixed regularization term and the frequentist fglasso performed almost identically in terms of edge selection, as would be expected, with the exception that the Bayesian approach allows access to the full posterior distribution so that any quantity of interest can be obtained, not just the posterior mode. The hierarchical Bayesian extension of the fglasso and the functional graphical horseshoe were directly compared to each other. The simulation results demonstrated that the functional graphical horseshoe is much better at avoiding false positives than either the frequentist or Bayesian fglasso and is still able to detect relatively weak signals. The superior balance between false positives and false negatives results in the functional graphical horseshoe exhibiting generally superior similarity to the underlying true graph. We also applied the functional graphical horseshoe to two applications in neuroimaging, one a previously studied EEG example, and the other involving DTI measurements to compare white matter integrity in people with and without a history of traumatic brain injury. 

The ability of the functional graphical horseshoe to avoid false positives is a critical characteristic for application in functional neuroimaging, as this field is often criticized for its abundance of false positives \citep{EklundEtAl16}. In addition, both EEG and MRI are notorious for having weak signal-to-noise ratios, making methods that are powerful at detecting weak signals quite useful. Indeed, the results in Section \ref{sec:ADNI} suggest that our proposed methods may be helpful for understanding functional reorganization, a process in which many weak connections form.

While our work here suggests promising results, especially with extending the graphical horseshoe to functional graphical models, much work still remains. The sensitivity of results to the selected basis remains to be explored. Our proposed Gibbs samplers are efficient in small to moderately-sized scenarios. One future line of research would be to explore more computationally efficient MCMC techniques, or other posterior approximation methods, appropriate for extremely high dimensional problems. There also remains deeper theoretical development of Bayesian approaches to functional graphical models, which to date is very limited.

\vskip 0.2in
\bibliographystyle{chicago}
\bibliography{wileyNJD-AMA}

\begin{thebibliography}{}

\bibitem[\protect\citeauthoryear{Andersson and Sotiropoulos}{Andersson and
  Sotiropoulos}{2016}]{andersson2016integrated}
Andersson, J.~L. and S.~N. Sotiropoulos (2016).
\newblock An integrated approach to correction for off-resonance effects and
  subject movement in diffusion mr imaging.
\newblock {\em Neuroimage\/}~{\em 125}, 1063--1078.

\bibitem[\protect\citeauthoryear{Banerjee, El~Ghaoui, and d'Aspremont}{Banerjee
  et~al.}{2008}]{BanerjeeEtAl08}
Banerjee, O., L.~El~Ghaoui, and A.~d'Aspremont (2008).
\newblock Model selection through sparse maximum likelihood estimation for
  multivariate gaussian or binary data.
\newblock {\em Journal of Machine Learning Research\/}~{\em 9}, 485--516.

\bibitem[\protect\citeauthoryear{Banerjee and Ghosal}{Banerjee and
  Ghosal}{2014}]{BanerjeeGhosal14}
Banerjee, S. and S.~Ghosal (2014).
\newblock Posterior convergence rates for estimating large precision matrices
  using graphical models.
\newblock {\em Electronic Journal of Statistics\/}~{\em 8\/}(2), 2111--2137.

\bibitem[\protect\citeauthoryear{Bosq}{Bosq}{2012}]{bosq2012linear}
Bosq, D. (2012).
\newblock {\em Linear processes in function spaces: theory and applications},
  Volume 149.
\newblock Springer Science \& Business Media.

\bibitem[\protect\citeauthoryear{Cabral, L., and Deco}{Cabral
  et~al.}{2014}]{CabralEtAl14}
Cabral, J., K.~M. L., and G.~Deco (2014).
\newblock Exploring the network dynamics underlying brain activity during rest.
\newblock {\em Progress in Neurobiology\/}~{\em 114}, 102--131.

\bibitem[\protect\citeauthoryear{Calhoun, Miller, Pearlson, and Adali}{Calhoun
  et~al.}{2014}]{CalhounEtal14}
Calhoun, V.~D., R.~Miller, G.~Pearlson, and T.~Adali (2014).
\newblock The chronnectome: Time-varying connectivity networks as the next
  frontier in fmri data discovery.
\newblock {\em Neuron\/}~{\em 84}, 262--274.

\bibitem[\protect\citeauthoryear{Carroll, Gajardo, Chen, Dai, Fan,
  Hadjipantelis, Han, Ji, M{\"u}ller, and Wang}{Carroll
  et~al.}{2020}]{carroll2020fdapace}
Carroll, C., A.~Gajardo, Y.~Chen, X.~Dai, J.~Fan, P.~Hadjipantelis, K.~Han,
  H.~Ji, H.~M{\"u}ller, and J.~Wang (2020).
\newblock fdapace: Functional data analysis and empirical dynamics.
\newblock {\em R package version 0.5\/}~{\em 5}.

\bibitem[\protect\citeauthoryear{Carroll, Gajardo, Chen, Dai, Fan,
  Hadjipantelis, Han, Ji, Mueller, and Wang}{Carroll et~al.}{2020}]{fdapacePkg}
Carroll, C., A.~Gajardo, Y.~Chen, X.~Dai, J.~Fan, P.~Z. Hadjipantelis, K.~Han,
  H.~Ji, H.-G. Mueller, and J.-L. Wang (2020).
\newblock {\em fdapace: Functional Data Analysis and Empirical Dynamics}.
\newblock R package version 0.5.5.

\bibitem[\protect\citeauthoryear{Carvalho, Polson, and Scott}{Carvalho
  et~al.}{2010}]{carvalho2010horseshoe}
Carvalho, C.~M., N.~G. Polson, and J.~G. Scott (2010).
\newblock The horseshoe estimator for sparse signals.
\newblock {\em Biometrika\/}~{\em 97\/}(2), 465--480.

\bibitem[\protect\citeauthoryear{Dawid and Lauritzen}{Dawid and
  Lauritzen}{1993}]{DawidLauritzen93}
Dawid, A.~P. and S.~L. Lauritzen (1993).
\newblock Hyper markov laws in the statistical analysis of decomposable
  graphical models.
\newblock {\em Annals of Statistics\/}~{\em 21\/}(3), 1272--1317.

\bibitem[\protect\citeauthoryear{Dempster}{Dempster}{1972}]{Dempster72}
Dempster, A.~P. (1972).
\newblock Covariance selection.
\newblock {\em Biometrika\/}~{\em 32}, 95--108.

\bibitem[\protect\citeauthoryear{Eklund, Nichols, and Knutsson}{Eklund
  et~al.}{2016}]{EklundEtAl16}
Eklund, A., T.~E. Nichols, and H.~Knutsson (2016).
\newblock {Cluster failure: Why fMRI inferences for spatial extent have
  inflated false-positive rates}.
\newblock {\em Proceedings of the National Academy of Sciences\/}~{\em
  113\/}(28), 7900--7905.

\bibitem[\protect\citeauthoryear{Friedman, Hastie, and Tibshirani}{Friedman
  et~al.}{2008}]{friedman2008sparse}
Friedman, J., T.~Hastie, and R.~Tibshirani (2008).
\newblock Sparse inverse covariance estimation with the graphical lasso.
\newblock {\em Biostatistics\/}~{\em 9\/}(3), 432--441.

\bibitem[\protect\citeauthoryear{Gelfand and Smith}{Gelfand and
  Smith}{1990}]{GelfandSmith90}
Gelfand, A.~E. and A.~F. Smith (1990).
\newblock {Sampling-based approaches to calculating marginal densities}.
\newblock {\em Journal of the American Statistical Association\/}~{\em
  85\/}(410), 398--409.

\bibitem[\protect\citeauthoryear{Greenlaw, Szefer, Graham, Lesperance, Nathoo,
  and Initiative}{Greenlaw et~al.}{2017}]{GreenlawEtAl17}
Greenlaw, K., E.~Szefer, J.~Graham, M.~Lesperance, F.~S. Nathoo, and A.~D.~N.
  Initiative (2017).
\newblock A bayesian group sparse multi-task regression model for imaging
  genetics.
\newblock {\em Bioinformatics\/}~{\em 33\/}(16), 2513--2522.

\bibitem[\protect\citeauthoryear{Helwig}{Helwig}{2018}]{helwig2018eegkit}
Helwig, N. (2018).
\newblock eegkit: Toolkit for electroencephalography data.
\newblock {\em URL: http://CRAN. R-project. org/package= eegkit. R package
  version\/}, 1--0.

\bibitem[\protect\citeauthoryear{Jenkinson, Bannister, Brady, and
  Smith}{Jenkinson et~al.}{2002}]{jenkinson2002improved}
Jenkinson, M., P.~Bannister, M.~Brady, and S.~Smith (2002).
\newblock Improved optimization for the robust and accurate linear registration
  and motion correction of brain images.
\newblock {\em Neuroimage\/}~{\em 17\/}(2), 825--841.

\bibitem[\protect\citeauthoryear{Jenkinson and Smith}{Jenkinson and
  Smith}{2001}]{jenkinson2001global}
Jenkinson, M. and S.~Smith (2001).
\newblock A global optimisation method for robust affine registration of brain
  images.
\newblock {\em Medical image analysis\/}~{\em 5\/}(2), 143--156.

\bibitem[\protect\citeauthoryear{Jolliffe}{Jolliffe}{2002}]{Jolliffe02}
Jolliffe, I. (2002).
\newblock {\em {Principal Component Analysis}\/} (2nd ed.).
\newblock New York: Springer.

\bibitem[\protect\citeauthoryear{Karhunen}{Karhunen}{1946}]{Kar46}
Karhunen, K. (1946).
\newblock {Zur Spektraltheorie Stochastischer Prozesse}.
\newblock {\em Annales Academi{\ae} Scientiarum Fennic{\ae}\/}~{\em 34}, 1--7.

\bibitem[\protect\citeauthoryear{Kook, Vaughn, DeMaster, Ewing-Cobbs, and
  Vannucci}{Kook et~al.}{2021}]{KookEtAl20}
Kook, J.~H., K.~A. Vaughn, D.~M. DeMaster, L.~Ewing-Cobbs, and M.~Vannucci
  (2021).
\newblock Bvar-connect: A variational bayes approach to multi-subject vector
  autoregressive models for inference on brain connectivity networks.
\newblock {\em Neuroinformatics\/}~{\em 19}, 39--56.

\bibitem[\protect\citeauthoryear{Kyung, Gill, Ghosh, Casella, et~al.}{Kyung
  et~al.}{2010}]{kyung2010penalized}
Kyung, M., J.~Gill, M.~Ghosh, G.~Casella, et~al. (2010).
\newblock Penalized regression, standard errors, and bayesian lassos.
\newblock {\em Bayesian Analysis\/}~{\em 5\/}(2), 369--411.

\bibitem[\protect\citeauthoryear{Lam and Fan}{Lam and
  Fan}{2009}]{lam2009sparsistency}
Lam, C. and J.~Fan (2009).
\newblock Sparsistency and rates of convergence in large covariance matrix
  estimation.
\newblock {\em Annals of statistics\/}~{\em 37\/}(6B), 4254.

\bibitem[\protect\citeauthoryear{Li and Solea}{Li and
  Solea}{2018}]{li2018nonparametric}
Li, B. and E.~Solea (2018).
\newblock A nonparametric graphical model for functional data with application
  to brain networks based on fmri.
\newblock {\em Journal of the American Statistical Association\/}~{\em
  113\/}(524), 1637--1655.

\bibitem[\protect\citeauthoryear{Li, Craig, and Bhadra}{Li
  et~al.}{2019}]{li2019graphical}
Li, Y., B.~A. Craig, and A.~Bhadra (2019).
\newblock The graphical horseshoe estimator for inverse covariance matrices.
\newblock {\em Journal of Computational and Graphical Statistics\/}, 1--24.

\bibitem[\protect\citeauthoryear{Lindow, Domin, Grothe, Horn, Eickhoff, and
  Lotze}{Lindow et~al.}{2016}]{LindowEtAl16}
Lindow, J., M.~Domin, M.~Grothe, U.~Horn, S.~B. Eickhoff, and M.~Lotze (2016).
\newblock Connectivity-based predictions of hand motor outcome for patients at
  the subacute stage after stroke.
\newblock {\em Frontiers in human neuroscience\/}~{\em 10}, 101.

\bibitem[\protect\citeauthoryear{Loeve}{Loeve}{1963}]{Loe46}
Loeve, M. (1963).
\newblock {\em Probability Theory. Van Nostrand}.

\bibitem[\protect\citeauthoryear{Makalic and Schmidt}{Makalic and
  Schmidt}{2016}]{makalic2015simple}
Makalic, E. and D.~F. Schmidt (2016).
\newblock A simple sampler for the horseshoe estimator.
\newblock {\em IEEE Signal Processing Letters\/}~{\em 23}, 179--182.

\bibitem[\protect\citeauthoryear{Meinshausen, B{\"u}hlmann, et~al.}{Meinshausen
  et~al.}{2006}]{meinshausen2006high}
Meinshausen, N., P.~B{\"u}hlmann, et~al. (2006).
\newblock High-dimensional graphs and variable selection with the lasso.
\newblock {\em The annals of statistics\/}~{\em 34\/}(3), 1436--1462.

\bibitem[\protect\citeauthoryear{Morris, Baladandayuthapani, Herrick, Sanna,
  and Gutstein}{Morris et~al.}{2011}]{morris2011automated}
Morris, J.~S., V.~Baladandayuthapani, R.~C. Herrick, P.~Sanna, and H.~Gutstein
  (2011).
\newblock Automated analysis of quantitative image data using isomorphic
  functional mixed models, with application to proteomics data.
\newblock {\em The annals of applied statistics\/}~{\em 5\/}(2A), 894.

\bibitem[\protect\citeauthoryear{Qiao, Guo, and James}{Qiao
  et~al.}{2019}]{qiao2019functional}
Qiao, X., S.~Guo, and G.~M. James (2019).
\newblock Functional graphical models.
\newblock {\em Journal of the American Statistical Association\/}~{\em
  114\/}(525), 211--222.

\bibitem[\protect\citeauthoryear{Qiao, Qian, James, and Guo}{Qiao
  et~al.}{2020}]{qiao2020doubly}
Qiao, X., C.~Qian, G.~M. James, and S.~Guo (2020).
\newblock Doubly functional graphical models in high dimensions.
\newblock {\em Biometrika\/}~{\em 107\/}(2), 415--431.

\bibitem[\protect\citeauthoryear{Rajaratnam, Massam, and Carvalho}{Rajaratnam
  et~al.}{2008}]{RajarEtAl08}
Rajaratnam, B., H.~Massam, and C.~M. Carvalho (2008).
\newblock Flexible covariance estimation in graphical gaussian models.
\newblock {\em The Annals of Statistics\/}~{\em 36\/}(6), 2818--2849.

\bibitem[\protect\citeauthoryear{Rue and Held}{Rue and Held}{2005}]{RueHeld05}
Rue, H. and L.~Held (2005).
\newblock {\em Gaussian Markov random fields: theory and applications}.
\newblock CRC press.

\bibitem[\protect\citeauthoryear{Rutgers, Toulgoat, Cazejust, Fillard,
  Lasjaunias, and Ducreux}{Rutgers et~al.}{2008}]{RutgersEtAl08}
Rutgers, D., F.~Toulgoat, J.~Cazejust, P.~Fillard, P.~Lasjaunias, and
  D.~Ducreux (2008).
\newblock White matter abnormalities in mild traumatic brain injury: a
  diffusion tensor imaging study.
\newblock {\em American Journal of Neuroradiology\/}~{\em 29\/}(3), 514--519.

\bibitem[\protect\citeauthoryear{Shappell, Caffo, Pekar, and
  Lindquist}{Shappell et~al.}{2019}]{ShappelEtAl19}
Shappell, H., B.~S. Caffo, J.~J. Pekar, and M.~A. Lindquist (2019).
\newblock Improved state change estimation in dynamic functional connectivity
  using hidden semi-markov models.
\newblock {\em NeuroImage\/}~{\em 191}, 243--257.

\bibitem[\protect\citeauthoryear{Smith}{Smith}{2002}]{smith2002fast}
Smith, S.~M. (2002).
\newblock Fast robust automated brain extraction.
\newblock {\em Human brain mapping\/}~{\em 17\/}(3), 143--155.

\bibitem[\protect\citeauthoryear{Solea and Li}{Solea and
  Li}{2020}]{solea2020copula}
Solea, E. and B.~Li (2020).
\newblock Copula gaussian graphical models for functional data.
\newblock {\em Journal of the American Statistical Association\/}, 1--13.

\bibitem[\protect\citeauthoryear{Storey}{Storey}{2003}]{storey2003positive}
Storey, J.~D. (2003).
\newblock The positive false discovery rate: a bayesian interpretation and the
  q-value.
\newblock {\em The Annals of Statistics\/}~{\em 31\/}(6), 2013--2035.

\bibitem[\protect\citeauthoryear{Van Der~Pas, Kleijn, and Van Der~Vaart}{Van
  Der~Pas et~al.}{2014}]{VanDerPasEtAl14}
Van Der~Pas, S.~L., B.~J. Kleijn, and A.~W. Van Der~Vaart (2014).
\newblock The horseshoe estimator: Posterior concentration around nearly black
  vectors.
\newblock {\em Electronic Journal of Statistics\/}~{\em 8\/}(2), 2585--2618.

\bibitem[\protect\citeauthoryear{Vandenberghe, Boyd, and Wu}{Vandenberghe
  et~al.}{1998}]{VanEtAl98}
Vandenberghe, L., S.~Boyd, and S.-P. Wu (1998).
\newblock Determinant maximization with linear matrix inequality constraints.
\newblock {\em SIAM journal on matrix analysis and applications\/}~{\em
  19\/}(2), 499--533.

\bibitem[\protect\citeauthoryear{Wang}{Wang}{2012}]{wang2012bayesian}
Wang, H. (2012).
\newblock Bayesian graphical lasso models and efficient posterior computation.
\newblock {\em Bayesian Analysis\/}~{\em 7\/}(4), 867--886.

\bibitem[\protect\citeauthoryear{Warnick, Guindani, Erhardt, Allen, Calhoun,
  and Vannucci}{Warnick et~al.}{2018}]{warnick2018bayesian}
Warnick, R., M.~Guindani, E.~Erhardt, E.~Allen, V.~Calhoun, and M.~Vannucci
  (2018).
\newblock A bayesian approach for estimating dynamic functional network
  connectivity in fmri data.
\newblock {\em Journal of the American Statistical Association\/}~{\em
  113\/}(521), 134--151.

\bibitem[\protect\citeauthoryear{Wasserman and Roeder}{Wasserman and
  Roeder}{2009}]{wasserman2009high}
Wasserman, L. and K.~Roeder (2009).
\newblock High dimensional variable selection.
\newblock {\em Annals of statistics\/}~{\em 37\/}(5A), 2178.

\bibitem[\protect\citeauthoryear{Xiang, Khare, and Ghosh}{Xiang
  et~al.}{2015}]{XiangEtAl15}
Xiang, R., K.~Khare, and M.~Ghosh (2015).
\newblock High dimensional posterior convergence rates for decomposable
  graphical models.
\newblock {\em Electronic Journal of Statistics\/}~{\em 9\/}(2), 2828--2854.

\bibitem[\protect\citeauthoryear{Yao, M{\"u}ller, and Wang}{Yao
  et~al.}{2005}]{yao2005functional}
Yao, F., H.-G. M{\"u}ller, and J.-L. Wang (2005).
\newblock Functional data analysis for sparse longitudinal data.
\newblock {\em Journal of the American Statistical Association\/}~{\em
  100\/}(470), 577--590.

\bibitem[\protect\citeauthoryear{Yuan and Lin}{Yuan and
  Lin}{2006}]{yuan2006model}
Yuan, M. and Y.~Lin (2006).
\newblock Model selection and estimation in regression with grouped variables.
\newblock {\em Journal of the Royal Statistical Society: Series B (Statistical
  Methodology)\/}~{\em 68\/}(1), 49--67.

\bibitem[\protect\citeauthoryear{Yuan and Lin}{Yuan and
  Lin}{2007}]{yuan2007model}
Yuan, M. and Y.~Lin (2007).
\newblock {Model selection and estimation in the Gaussian graphical model}.
\newblock {\em Biometrika\/}~{\em 94\/}(1), 19--35.

\bibitem[\protect\citeauthoryear{Zapata, Oh, and Petersen}{Zapata
  et~al.}{2019}]{zapata2019partial}
Zapata, J., S.-Y. Oh, and A.~Petersen (2019).
\newblock Partial separability and functional graphical models for multivariate
  gaussian processes.
\newblock {\em arXiv preprint arXiv:1910.03134\/}.

\bibitem[\protect\citeauthoryear{Zhang, Baladandayuthapani, Neville, Quevedo,
  and Morris}{Zhang et~al.}{2021}]{Zhang_etal_21}
Zhang, L., V.~Baladandayuthapani, Q.~Neville, K.~Quevedo, and J.~S. Morris
  (2021).
\newblock Bayesian functional graphical models.
\newblock arXiv preprint arXiv:2108.05034.

\bibitem[\protect\citeauthoryear{Zhang, Begleiter, Porjesz, Wang, and
  Litke}{Zhang et~al.}{1995}]{zhang1995event}
Zhang, X.~L., H.~Begleiter, B.~Porjesz, W.~Wang, and A.~Litke (1995).
\newblock Event related potentials during object recognition tasks.
\newblock {\em Brain Research Bulletin\/}~{\em 38\/}(6), 531--538.

\bibitem[\protect\citeauthoryear{Zhu, Strawn, and Dunson}{Zhu
  et~al.}{2016}]{zhu2016bayesian}
Zhu, H., N.~Strawn, and D.~B. Dunson (2016).
\newblock Bayesian graphical models for multivariate functional data.
\newblock {\em The Journal of Machine Learning Research\/}~{\em 17\/}(1),
  7157--7183.

\bibitem[\protect\citeauthoryear{Zhu, Shen, and Pan}{Zhu
  et~al.}{2014}]{zhu2014structural}
Zhu, Y., X.~Shen, and W.~Pan (2014).
\newblock Structural pursuit over multiple undirected graphs.
\newblock {\em Journal of the American Statistical Association\/}~{\em
  109\/}(508), 1683--1696.

\end{thebibliography}

\end{document}